\documentclass[prd,preprint,aps,showpacs,tightenlines,superscriptaddress]{revtex4-1}
\usepackage[dvips]{graphicx}
\usepackage{dcolumn}
\usepackage{epsfig}
\usepackage{amssymb}

\newcommand{\ket}[1]{|{#1}\rangle}
\newcommand{\bra}[1]{\langle{#1}|}
\newcommand{\inp}[2]{\langle{#1}|{#2}\rangle}
\def\slash#1{\not\!#1}

\begin{document}

\title{Pion-Exchange and Fermi-Motion Effects on the Proton-Deuteron Drell-Yan Process}
\author{H. Kamano}
\affiliation{Research Center for Nuclear Physics, Osaka University, Ibaraki, Osaka 567-0047, Japan}
\author{T.-S. H. Lee}
\affiliation{Physics Division, Argonne National Laboratory, Argonne, Illinois 60439, USA}

\begin{abstract}
Within a nuclear model that the deuteron has $NN$ and $\pi NN$ components,
we derive convolution formula for investigating 
the Drell-Yan process in proton-deuteron ($pd$) reactions.
The contribution from the $\pi NN$ component is expressed in terms of 
a pion momentum distribution that depends sensitively on the $\pi NN$ form factor.
With a $\pi NN$ form factor determined by fitting the $\pi N$ scattering 
data up to invariant mass $W=$ 1.3 GeV, 
we find that the pion-exchange and nucleon Fermi-motion effects  
can change significantly the ratios between the proton-deuteron 
and proton-proton Drell-Yan cross sections, $R_{pd/pp}=\sigma^{pd}/(2\sigma^{pp})$,
in the region where the partons emitted from the target deuteron are in the Bjorken 
$x_2 \gtrsim 0.4$ region.
The calculated ratios $R_{pd/pp}$ at 800 GeV agree with the available data.
Predictions at 120 GeV for analyzing the forthcoming data from Fermilab are presented.
\end{abstract}


\maketitle

\section{Introduction}
\label{sec:int}

Since the asymmetry between the anti-up ($\bar u$)
and anti-down ($\bar d$) quark distributions in the proton was revealed
by the New Muon Collaboration~\cite{nmc} (NMC), a series of
experiments~\cite{peng1992,peng1998,peng1998a,peng2001} on the
di-muons ($\mu^+\mu^-)$ production from the Drell-Yan~\cite{dy} (DY) processes in
$pp$ and $pd$ collisions had been performed at 
Fermi National Accelerator Laboratory (Fermilab).
The objective was to extract the $\bar{d}/\bar{u}$
ratio of the parton distribution functions (PDFs) in the proton.
The information from these experiments and the measurements~\cite{nmc,na51,hermes}
of deep inelastic scattering (DIS) of leptons from the nucleon have confirmed 
the NMC's finding, $\bar{d}/\bar{u} > 1 $, only in the region of low Bjorken 
$x \lesssim 0.35$.

The ratio $\bar{d}/\bar{u} > 1$ signals the nonperturbative nature of the sea of the proton.
Its dynamical origins have been investigated~\cite{thom83,thom98,kuma98,spet97,
kuma91,hwan93,szcz94,koep96,niko99,albe00,eich93,szcz96,poby99,doro93,meln99}
rather extensively. Precise experimental determination of $\bar{d}/\bar{u}$ for higher 
$x > 0.35$ is needed to distinguish more decisively these models and to develop 
a deeper understanding of the the sea of the proton.
This information will soon become available from a forthcoming 
experiment~\cite{peng2011} at Fermilab.

In analyzing the DY data on the deuteron~\cite{peng1992,peng1998,peng1998a,peng2001} 
and nuclei~\cite{dynucl-1990,dynucl-1999,dypi-1989,holt2005}, it is common to neglect 
the nuclear effects that are known to be important in analyzing the DIS data.
It is well established that the nuclear effect due to the nucleon Fermi motion (FM) 
can influence significantly the DIS cross sections, in particular in the large $x$ region.
It is also known that the contributions from the virtual pions in nuclei must be 
considered for a quantitative understanding of the parton distributions in nuclear medium.
Thus it is interesting and also important to develop an approach to investigate 
these two nuclear effects on the $pd$ DY process. 
This is the main purpose of this work.
We will  apply our formula to analyze the available data at 800 GeV~\cite{peng2001}
and make predictions for the forthcoming experiment~\cite{peng2011}.

It is instructive to describe here how the DY data were analyzed, as
described in, for example, Ref.~\cite{peng2001}.
The leading-order DY cross sections from $pN$ collision with 
$N= p$ (proton), $n$ (neutron) is written as
\begin{eqnarray}
\frac{d\sigma^{pN}}{dx_1 dx_2} 
&=&
\frac{4\pi \alpha^2}{9 M^2} \sum_{q} \hat e^2_q
[f^q_p(x_1) f^{\bar{q}}_N(x_2)+f^{\bar{q}}_p(x_1) f^{q}_N(x_2)] \,,
\label{eq:dy-exp}
\end{eqnarray}
where the sum is over all quark flavors, $\hat e_q$ is the quark
charge, $f^q_N(x)$ is the parton distribution of parton $q$ in hadron
$N$, and $M$ is the virtual photon or di-lepton mass.
Here $x_1$ and $x_2$ are the Bjorken-$x$ of partons from the beam ($p$)
and target ($N$), respectively 
(see Sec.~\ref{sec:numerical-1} for explicit definitions of $x_1$ and $x_2$).
The DY cross section for $pd$ is taken to be
\begin{eqnarray}
\frac{d\sigma^{pd}}{dx_1 dx_2} 
&=& 
\frac{d\sigma^{pp}}{dx_1 dx_2} + \frac{d\sigma^{pn}}{dx_1 dx_2}\,.
\label{eq:dy-pd-exp}
\end{eqnarray}
Obviously, Eq.~(\ref{eq:dy-pd-exp}) does not account for the nucleon 
Fermi-motion and pion-exchange effects.
To make progress, it is necessary to investigate under what assumptions 
Eqs.~(\ref{eq:dy-exp}) and~(\ref{eq:dy-pd-exp}) can be derived from 
a formulation within which these two nuclear effects
are properly accounted for.

We start with a nuclear model within which the deuteron wave function 
has $NN$ and $\pi NN$ components.
Such a model was developed in the study of $\pi NN$ system~\cite{garcilazo}.
We will derive convolution formula to express the DY cross section in terms of 
the momentum distributions $\rho (\vec{p})$  calculated from the $NN$ component
and $\rho (\vec{k}_\pi)$ from $\pi NN$ component.
Since $\pi NN$ component is much weaker, it is a good approximation to use
the $NN$ component generated from the available realistic $NN$ potentials~\cite{v14}.
In the same leading order approximation, the resulting
$\rho (\vec{k}_\pi)$ depend sensitively on the $\pi NN$ form factor.
An essential feature of our approach is to determine this form factor
from fitting the $\pi N$ scattering data. 
This provides an empirical constraint on our predictions of the pion effects 
on the proton-deuteron DY cross sections in the un-explored large $x$ region.

To see clearly the content of our approach, we will give a rather elementary 
derivation of our formula with all approximations specified explicitly.
In Sec.~\ref{sec:DY-formula}, we start with the general covariant form of the DY cross 
section and indicate the procedures needed to obtain  the well known
 $q\bar{q} \to \mu^+\mu^-$ cross section $\sigma^{q\bar{q}}$.
The same procedures are then used to derive  formula for
calculating the $pp$ and $pn$ DY cross sections from
$\sigma^{q\bar{q}}$ and a properly defined  
PDFs $f^{q}_N$ of the nucleon.

In Sec.~\ref{sec:pdcs}, we use the impulse approximation to derive
the formula for calculating $pd$ DY cross sections from
$\sigma^{q\bar{q}}$, $f^{q}_N$, and the momentum distributions
$\rho_N(p)$ for nucleon and $\rho_\pi (k)$ for pions
in the deuteron. The calculations of these two momentum distributions within 
the considered $\pi NN$ model are explained in Sec.~\ref{sec:dist}.

In Sec.~\ref{sec:numerical}, we develop the procedures for applying the developed formula
to perform  numerical calculations of $pp$ and $pd$ DY cross sections
using the available PDFs~\cite{lai97,mart96,gluc95,plot95,ceteq5m} and
realistic deuteron wave functions~\cite{v14}. 
In Sec.~\ref{sec:results}, we present  results to
compare with the available data at 800 GeV~\cite{peng2001} 
and make predictions for analyzing the forthcoming experiment~\cite{peng2011}.
A summary is given in Sec.~\ref{sec:summary}.

\section{Formula for DY Cross sections}
\label{sec:DY-formula}

The formula presented in this section are derived from using
the  Bjorken-Drell~\cite{bd-book} conventions for the
Dirac matrices and the field operators for leptons, nucleons, pions, and photons.
We  choose the normalization that
the plane-wave state
$\ket{\vec{k}}$ is normalized as
$\inp{\vec{k}}{\vec{k}^{\,'}} = \delta(\vec{k}-\vec{k}^{\,'})$ and the bound
states $\ket{\Phi_\alpha}$ of composite particles, nucleons or nuclei,
are normalized as $\inp{\Phi_\alpha}{\Phi_\beta} =\delta_{\alpha,\beta}$.
To simplify the presentation, spin indices are suppressed;
i.e. $\ket{\vec{k}_a}$ represents $\ket{\vec{k}_a,\lambda_a}$ for a particle
$a$ with helicity $\lambda_a$.
Thus the formula presented here are only for the spin averaged cross sections
which are our focus in this paper.

\begin{figure}[t]
\includegraphics[clip,width=0.5\textwidth]{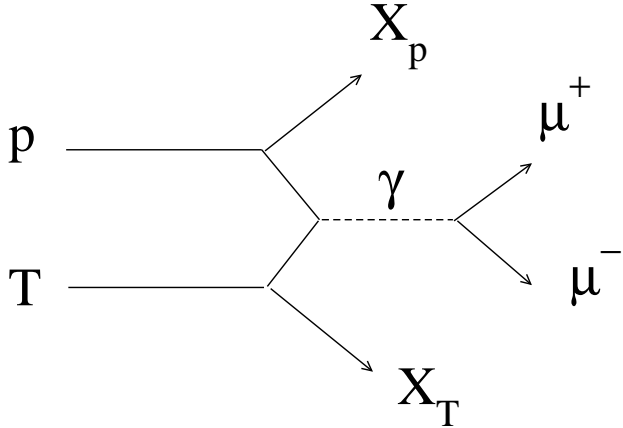}
\caption{DY process.}
\label{fig:dy-mech}
\end{figure}

We consider the di-muons production from the DY processes of  
hadron ($h$)-hadron ($T$) collisions:
\begin{eqnarray}
h(p_h) + T(p_T) &\to& \mu^+(k_+) + \mu^-(k_{-})+X_h(p_{X_h}) + X_T(p_{X_T}) ,
\label{eq:dy-process}
\end{eqnarray}
where $X_h$ and $X_T$ are the  undetected fragments, and the four-momentum 
of each particle is given within the parenthesis. 
In terms of the partonic $q   \bar{q}\to \gamma \to \mu^+ \mu^-$ mechanism,
illustrated in Fig.~\ref{fig:dy-mech}, the covariant form of the di-muons production 
cross section can be written as 
\begin{eqnarray}
d\sigma 
&=& 
\frac{(2\pi)^4}{4[(p_h\cdot p_T)^2-m^2_hm^2_T]^{1/2}}
\frac{1}{(2\pi)^6}\frac{d\vec{k}_+}{2E_+}\frac{d\vec{k}_-}{2E_-}
\frac{1}{q^4}f^{\mu\nu}(k_+,k_-)F_{\mu\nu}(p_h,p_T,q) \,, 
\label{eq:dy-crst}
\end{eqnarray}
where $m_h$ and $m_T$ are the masses for $h$ and $T$, 
respectively, $ E_{\pm}=\sqrt{\vec{k}^{\,2}_{\pm} + m^2_\mu}$ are the energies of 
muons $\mu^{\pm}$, and $q = k_+ + k_- $ is the momentum of the virtual photon.
The leptonic tensor is defined by
\begin{eqnarray}
f^{\mu\nu}(k_+,k_-) 
&=&  
(2\pi)^6 (2E_+)(2E_-)
\bra{\vec{k}_+\vec{k}_-} j^\mu(0) \ket{0} \bra{0} j^\nu(0) \ket{\vec{k}_+\vec{k}_-}\,.
\label{eq:f-lepton}
\end{eqnarray}
Here taking summation of lepton spins is implied.
The leptonic current is
\begin{eqnarray}
j^\mu(x) &=& e\bar{\psi}_\mu(x)\gamma^\mu \psi_\mu(x)\,,
\label{eq:c-lepton}
\end{eqnarray}
where $\psi_\mu(x)$ is the field operator for muon, and
$e=\sqrt{4\pi\alpha}$ with $\alpha=1/137$.
By using the definitions Eqs.~(\ref{eq:f-lepton}) and~(\ref{eq:c-lepton}), 
it is straightforward to get the following analytic form of the lepton tensor
\begin{eqnarray}
f^{\mu\nu}(k_+,k_-)
&=& 
-4e^2[k^\mu_+ k^\nu_- +k^\nu_+ k^\mu -g^{\mu\nu}(k_+\cdot k_- +m^2_\mu)] .
\label{eq:lepton-t}
\end{eqnarray}

Within the parton model, the hadronic tensor in Eq.~(\ref{eq:dy-crst}) is determined by  
the current $J_\mu(x)$ carried by partons $q$ or $\bar{q}$
\begin{eqnarray}
F_{\mu\nu}(p_h,p_T,q)
&=&
\sum_{X_h,X_T} (2\pi)^6(2E_h)(2E_T) 
\int d\vec{p}_{X_h}d\vec{p}_{X_T}
\delta^4(p_h+p_T-p_{X_h}-p_{X_T}-q)
\nonumber \\
&&\times
\bra{p_h p_T} J_\mu(0) \ket{\vec{p}_{X_h}d\vec{p}_{X_T}}
\bra{\vec{p}_{X_T}\vec{p}_{X_h}} J_\nu(0) \ket{p_hp_T} .
\label{eq:f-hadron}
\end{eqnarray}
Here it is noted that, throughout this paper, we shall take the summation (average) 
for the spins of final (initial) particles appearing in hadronic tensors.
\begin{eqnarray}
J_\mu(x)=\sum_{q} \hat{e}_q e \bar{\psi}_{q}(x)\gamma^\mu \psi_q(x)\,,
\label{eq:c-parton}
\end{eqnarray}
where $\psi_q(x)$ is the field operator for a quark
$q$  with charge $\hat{e}_q e$; i.e $\hat{e}_u=2/3$
and $\hat{e}_d=-1/3$ for up and down quarks, respectively.

The above covariant expressions are convenient for deriving the formula that
can express the hadron-hadron  DY cross sections in terms of  the elementary partonic  
$q\bar{q} \to \mu^+\mu^-$ cross sections. 
To get such formula, we first show how the elementary 
$q\bar{q} \to \mu^+ \mu^-$ cross section can be derived
from Eq.~(\ref{eq:dy-crst}) with $h=q$ and $T=\bar{q}$. 
We then derive the formula for calculating proton-nucleon DY cross section.

\subsection{$q\bar{q} \to \mu^+\mu^-$ cross section}

Explicitly, Eq.~(\ref{eq:dy-crst}) for the
$q(p_q)+ \bar{q}(p_{\bar{q}}) \to \mu^+(k_+)+ \mu^-(k_-)$ process is
\begin{eqnarray}
d\sigma^{q\bar{q}} 
&=&
\frac{(2\pi)^4}{4[(p_q\cdot p_{\bar{q}})^2-m^4_q]^{1/2}}
\left\{
\frac{1}{(2\pi)^6}
\frac{d\vec{k}_+}{2E_+}\frac{d\vec{k}_-}{2E_-}
\frac{1}{q^4}f^{\mu\nu}(k_+,k_-)F^{q\bar{q}}_{\mu\nu}(p_q,p_{\bar{q}},q)
\right\}
\,.
\label{eq:dy-qbarq-crst}
\end{eqnarray}
The next step is to replace the intermediate states $\ket{\vec{p}_{X_h} \vec{p}_{X_T}}$ by the 
the vacuum state $\ket{0}$ in evaluating the hadronic tensor Eq.~(\ref{eq:f-hadron}). 
We thus have
\begin{eqnarray}
F^{q\bar{q}}_{\mu\nu}(p_q,p_{\bar{q}},q)
&=&
(2\pi)^6 (2E_q)(2E_{\bar{q}}) 
\bra{p_{\bar{q}}p_q} J_\mu(0) \ket{0}
\bra{0} J_\nu(0) \ket{p_q,p_{\bar{q}}}
\delta^4(p_q+p_{\bar{q}}-q)\,.
\nonumber\\
\label{eq:qq-tensor}
\end{eqnarray}
Substituting parton current~(\ref{eq:c-parton}) into Eq.~(\ref{eq:qq-tensor}),
the hadronic tensor $F^{q\bar{q}}_{\mu\nu}$ then has a form that is the same as the leptonic 
tensor $f^{\mu\nu}$ defined by Eqs.~(\ref{eq:f-lepton}) and~(\ref{eq:c-lepton})
except that the momentum variables and charges are different. 
By appropriately changing the momentum variables in Eq.~(\ref{eq:lepton-t}),
we obtain
\begin{eqnarray}
F^{q\bar{q}}_{\mu\nu}(p_q,p_{\bar{q}},q) 
&=& 
- (\hat{e}_qe)^2[p_q^\mu p_{\bar{q}}^\nu +  p_q^\nu p_{\bar{q}}^\mu 
-g^{\mu\nu}(p_q\cdot p_{\bar{q}} +m^2_q)]\delta^4(p_q+p_{\bar{q}}-q) \,.
\label{eq:parton-t}
\end{eqnarray}
Here, compared with the lepton tensor case [Eq.~(\ref{eq:lepton-t})],
the difference of factor $4$ is because 
the average of quark and anti-quark spins is taken for this case. 
By using Eqs.~(\ref{eq:lepton-t}) and~(\ref{eq:parton-t}), 
Eq.~(\ref{eq:dy-qbarq-crst}) for the cross sections of
$q(p_q)+ \bar{q}(p_{\bar{q}}) \to \mu^+(k_+)+ \mu^-(k_-)$
can then be written as
\begin{eqnarray}
d\sigma^{q\bar{q}}(p_q,p_{\bar{q}})
&=&
\frac{(2\pi)^4}{4[(p_q\cdot p_{\bar{q}})^2-m_q^4]^{1/2}}
\frac{1}{(2\pi)^6}\frac{d\vec{k}_+}{2E_+}\frac{d\vec{k}_-}{2E_-}
\frac{1}{q^4}\delta^4(p_q+p_{\bar{q}}-q) \nonumber \\
&& \times
8\hat e_q^2 e^4
\left[
k_+\cdot p_qk_-\cdot p_{\bar{q}} + k_-\cdot p_qk_+\cdot p_{\bar{q}}
+m^2_q\frac{(k_+{+}k_-)^2}{2} + m^2_\mu\frac{(p_q{+}p_{\bar{q}})^2}{2}
\right] .
\label{eq:crst-qq}
\end{eqnarray}

It is convenient to express the $q\bar{q}$ DY cross section  in terms of
the invariant function $q^2 = (p_q+p_{\bar{q}})^2 = (k_{+}+k_{-})^2$.
After some derivations and accounting for the color degrees of freedom of 
quarks, we arrive
\begin{eqnarray}
\frac{d\sigma^{q\bar{q}}(p_q,p_{\bar{q}})}{dq^2} 
&=&
\frac{4\pi \alpha^2}{q^2}\hat{e}_q^2
\frac{1}{3N_c} \frac{[q^2-{4m^2_\mu}]^{1/2}} {[q^2-{4m^2_q}]^{1/2}} 
{
\left(1+\frac{2m_\mu^2}{q^2}\right)
\left(1+\frac{2m_q^2}{q^2}\right)
}
\delta(q^2-(p_q+p_{\bar{q}})^2) \,,
\end{eqnarray}
where $N_c$ is the number of colors. Taking $N_c=3$ and 
considering  $q^2 \gg m^2_\mu $ and $q^2 \gg m^2_q$, we then obtain the familiar form
\begin{eqnarray}
\frac{d\sigma^{q\bar{q}}(p_q,p_{\bar{q}})}{dq^2} &=&
\frac{4\pi \alpha^2}{9q^2}\hat{e}_q^2
\delta \left(q^2-(p_q+p_{\bar{q}})^2 \right) .
\label{eq:dy-0}
\end{eqnarray}
The above expression is identical to the commonly used expression,
as given, for example, by the CETEQ group~\cite{ceteq5m}.

In Sec.~\ref{sec:dycs}, we will derive formula expressing
the $pN$  cross sections in terms of 
$d\sigma^{q\bar{q}}(p_q,p_{\bar{q}})/dq^2$ given in Eq.~(\ref{eq:dy-0}).

\subsection{$pN$ DY cross sections}
\label{sec:dycs}

To simplify the presentation, we only present formula for
$q$ in the projectile $p$ and $\bar{q}$ in the target $N$.
The term from the interchange $q\leftrightarrow \bar{q}$ will be included only
in the final expressions for calculations.

Equation~(\ref{eq:dy-crst}) for the 
$p(p_p)+ N(p_N)\to \mu^+(k_+)+\mu^-(k_-) + X_p(p_{X_p})+X_N(p_{X_N}) $ process  is
\begin{eqnarray}
d\sigma^{pN} &=&
\frac{(2\pi)^4}{4[(p_p \cdot p_N)^2 -m^2_pm^2_N]^{1/2}}
\left\{
\frac{1}{(2\pi)^6}\frac{d\vec{k}_+}{2E_+}\frac{d\vec{k}_-}{2E_-}
\frac{1}{q^4}f^{\mu\nu}(k_+,k_-)F^{pN}_{\mu\nu}(p_p,p_N,q)
\right\} ,
\label{eq:dy-crst-pn}
\end{eqnarray}
where the hadronic tensor, defined by  Eq.~(\ref{eq:f-hadron}), is
\begin{eqnarray}
F^{pN}_{\mu\nu}(p_p,p_N,q)
&=&
(2\pi)^6 (2E_p)(2E_N)
\sum_{X_p,X_N} \int d\vec{p}_{X_p} d\vec{p}_{X_N} \delta^4(p_p+p_N-p_{X_p}-p_{X_N}-q)
\nonumber \\
&&
\qquad\qquad\qquad\qquad\quad
\times 
\bra{ {\vec p_N} {\vec p_p} } J_\mu(0) \ket{ {\vec p_{X_p}} {\vec p_{X_N}} }
\bra{ {\vec p_{X_N}} {\vec p_{X_p}} } J_\nu(0) \ket{ {\vec p_p} {\vec p_N} } .
\label{eq:pN-tensor}
\end{eqnarray}
Within the parton model, the DY cross sections are calculated from the matrix element
$\bra{q\bar{q}} J_\mu(0) \ket{0}\bra{0} J_\nu(0) \ket{q\bar{q}}$  
which is  due to the annihilation of a $q$ ($\bar{q}$) from the projectile $p$ and 
a $\bar{q}$ ($q$) from the target $N$ into a photon. To identify such matrix elements, 
we insert a complete set of $q\bar{q}$ states (omitting spin indices) 
\begin{eqnarray}
1 = \int d \vec{p}_{q}d \vec{p}_{\bar{q}} 
\ket{\vec{p}_{q}\vec{p}_{\bar{q}}} \bra{\vec{p}_{\bar{q}}\vec{p}_{q}} ,
\nonumber
\end{eqnarray}
into Eq.~(\ref{eq:pN-tensor}). 
We then have
\begin{eqnarray}
F^{pN}_{\mu\nu}(p_p,p_N,q)
&=&
(2\pi)^6 (2E_p)(2E_N) 
\sum_{X_p,X_N} \int d\vec{p}_{X_p} d\vec{p}_{X_N} \delta^4(p_p+p_N-p_{X_p}-p_{X_N}-q) 
\nonumber \\
&& \times 
\int d\vec{p}_{q}d\vec{p}_{\bar{q}}
\int d\vec{p}^{\,'}_{q}d\vec{p}^{\,'}_{\bar{q}}
\inp{ \vec{p}_p }{ \vec{p}_q\vec{p}_{X_p} }
\inp{ \vec{p}_N }{ \vec{p}_{\bar{q}}\vec{p}_{X_N} }
\inp{ \vec{p}_{X_p}\vec{p}^{\,'}_q }{ \vec{p}_p }
\inp{ \vec{p}_{X_N}\vec{p}^{\,'}_{\bar{q}} }{ \vec{p}_N}
\nonumber \\
&& \times
\bra{ \vec{p}_{\bar{q}}\vec{p}_{q} } J_\mu(0) \ket{0}
\bra{0} J_\nu(0) \ket{\vec{p}^{\,'}_q \vec{p}^{\,'}_{\bar{q}}} . 
 \label{eq:pn-1}
\end{eqnarray}

By momentum conservation, the overlap functions in the above
equation can be written as
\begin{eqnarray}
\inp{ \vec{p}_q \vec{p}_{X} }{ \vec{p}_p } 
&=& 
\bra{ \vec{p}_X } b_{\vec{p}_q} \ket{ \vec{p}_p }
\nonumber \\
&=&
\phi_{\vec{p}_p}(\vec{p}_q ,\vec{p}_{X})
\delta(\vec{p}_p-\vec{p}_q -\vec{p}_{X}),
\label{eq:pn-2}
\end{eqnarray}
where $b_{\vec{p}_q}$ is the annihilation operator of quark $q$.

By using the definition ~(\ref{eq:pn-2}), Eq.~(\ref{eq:pn-1}) can be written as
\begin{eqnarray}
F^{pN}_{\mu\nu}(p_p,p_N,q)
&=&
(2\pi)^6(2E_p)(2E_N)
\sum_{X_P,X_N} \int d\vec{p}_{X_p} d\vec{p}_{X_N}
\delta^4(p_p+p_N-p_{X_p}-p_{X_N}-q)
\nonumber \\
&&
\times
\int d\vec{p}_q |\phi_{\vec{p}_p}(\vec{p}_q ,\vec{p}_{X_p})|^2
\delta(\vec{p}_p-\vec{p}_q -\vec{p}_{X_p})
\nonumber \\
&&
\times
\int d\vec{p}_{\bar{q}} | \phi_{\vec{p}_N}(\vec{p}_{\bar{q}} ,\vec{p}_{X_N})|^2
\delta(\vec{p}_N-\vec{p}_{\bar{q}} -\vec{p}_{X_N})
\nonumber \\
&&
\times
\bra{ \vec{p}_{\bar{q}}\vec{p}_{q} } J_\mu(0) \ket{0}
\bra{0} J_\nu(0) \ket{ \vec{p}_q \vec{p}_{\bar{q}} }  .
\label{eq:pn-3}
\end{eqnarray}

The evaluation of Eq.~(\ref{eq:pn-3}) needs rather detail information 
about the undetected fragments $X_p$ and $X_N$ because of the dependence of 
$\delta^4(p_p+p_N-p_{X_p}-p_{X_N}-q)$ on their energies $p^0_{X_p}$ and $p^0_{X_N}$.
To simplify the calculation, we follow the common practice to neglect the explicit 
dependence of the energy $p^0_{X_p}$ and $p^0_{X_N}$ of the undetected fragments. 
This amounts to using the approximation $p^0_{X_p}\sim \epsilon_1$ and
$p^0_{X_N}\sim \epsilon_2$, 
where $\epsilon_1$ and $\epsilon_2$ are some constant energies, 
to write  
\begin{eqnarray}
\delta^4(p_p+p_N-p_{X_p}-p_{X_N}-q)
&\sim &
\delta(\vec{p}_p+\vec{p}_N-\vec{p}_{X_p}-\vec{p}_{X_N}-\vec{q})
\delta(p^0_p+p^0_N - \epsilon_1-\epsilon_2 - q^0) .
\nonumber\\ 
\label{eq:delta-f}
\end{eqnarray}

We now define
\begin{eqnarray}
f^{q}_{\vec{p}_p}(\vec{p}_q) 
&=& 
\sum_{X_p} \int d\vec{p}_{X_p}
|\phi_{\vec{p}_p}(\vec{p}_q ,\vec{p}_{X_p})|^2
\delta(\vec{p}_p-\vec{p}_q -\vec{p}_{X_p}) ,
\label{eq:q-dis} 
\end{eqnarray}
for the projectile $p$, and 
\begin{eqnarray}
f^{\bar{q}}_{\vec{p}_N}(\vec{p}_{\bar{q}}) 
&=&
\sum_{X_N} \int d\vec{p}_{X_N}
|\phi_{\vec{p}_N}(\vec{p}_{\bar{q}} ,\vec{p}_{X_N})|^2
\delta(\vec{p}_N-\vec{p}_{\bar{q}} -\vec{p}_{X_N}) ,
\label{eq:barq-dis}
\end{eqnarray}
for the target $N$. 
These two definitions and the approximation~(\ref{eq:delta-f}) 
allow us to cast Eq.~(\ref{eq:pn-3}) into the following form
\begin{eqnarray}
F^{pN}_{\mu\nu}(p_p,p_N,q)
&=&
(2\pi)^6 (2E_p)(2E_N)\sum_{q}\int d\vec{p}_q d\vec{p}_{\bar{q}}
f^q_{\vec{p}_p} (\vec{p}_q) f^{\bar{q}}_{\vec{p}_N} (\vec{p}_{\bar{q}})
\nonumber \\
&&\times
\bra{ \vec{p}_{\bar{q}}\vec{p}_{q} } J_\mu(0) \ket{0} \bra{0} J_\nu(0)
\ket{ \vec{p}_q \vec{p}_{\bar{q}} }  
\nonumber \\
&&\times 
\delta(\vec{p}_q+\vec{p}_{\bar{q}}-\vec{q})
\delta(p^0_p+p^0_N - \epsilon_1-\epsilon_2 - q^0)  .
\label{eq:pn-4}
\end{eqnarray}

We next make a reasonable approximation that
$\epsilon_1$ ($\epsilon_2$) in Eq.~(\ref{eq:pn-4}) is the difference between
the energy of the projectile $p$ (target $N$) and the removed
parton $q$ ($\bar{q}$); namely assuming 
\begin{eqnarray}
\delta(p^0_p+p^0_N - \epsilon_1-\epsilon_2 - q^0)
&=&
\delta \left( (p^0_p-\epsilon_1)+(p^0_N-\epsilon_2)-q^0 \right) 
\nonumber \\
&\sim& \delta(p^0_q+p^0_{\bar{q}}-q^0)\,.
\label{eq:app-2}
\end{eqnarray}
Then Eq.~(\ref{eq:pn-4}) can be written as
\begin{eqnarray}
F^{pN}_{\mu\nu}(p_p,p_N,q)
&=&
\sum_{q}\int d\vec{p}_q d\vec{p}_{\bar{q}}
f^q_{\vec{p}_p} (\vec{p}_q) f^{\bar{q}}_{\vec{p}_N} (\vec{p}_{\bar{q}})
\frac{E_pE_N}{E_qE_{\bar{q}}} \nonumber \\
&&\times 
\{(2\pi)^6 (2E_q) (2E_{\bar{q}})
\bra{ \vec{p}_{\bar{q}}\vec{p}_{q} } J_\mu(0) \ket{0}
\bra{0} J_\nu(0) \ket{\vec{p}_q \vec{p}_{\bar{q}}} 
\delta^4(p_q+p_{\bar{q}}-q) \}\,. 
\nonumber \\
\label{eq:pn-5}
\end{eqnarray}
The quantity within the bracket $\{...\}$  in the above equation is just
the hadronic tensor $F^{q\bar q}_{\mu\nu}(p_q,p_{\bar{q}})$, defined 
in Eq.~(\ref{eq:qq-tensor}),
for $q\bar{q}$ system.
We thus have
\begin{eqnarray}
F^{pN}_{\mu\nu}(p_p,p_N,q)
&=&\sum_{q}\int d\vec{p}_q d\vec{p}_{\bar{q}} 
f^q_{\vec{p}_p} (\vec{p}_q) f^{\bar{q}}_{\vec{p}_N} (\vec{p}_{\bar{q}})
\frac{E_pE_N}{E_qE_{\bar{q}}}F^{q\bar q}_{\mu\nu}(p_q,p_{\bar{q}}) \,.
\label{eq:pN-tensor-1}
\end{eqnarray}

Substituting Eq.~(\ref{eq:pN-tensor-1}) into Eq.~(\ref{eq:dy-crst-pn}), we then have
\begin{eqnarray}
d\sigma^{pN}
&=&
\sum_{q}\int d\vec{p}_q d\vec{p}_{\bar{q}}
[f^q_{\vec{p}_p} (\vec{p}_q) f^{\bar{q}}_{\vec{p}_N} (\vec{p}_{\bar{q}})
]\frac{E_pE_N}{E_qE_{\bar{q}}}
\frac{(2\pi)^4}{4[(p_p\cdot p_N)^2-m^2_pm^2_N]^{1/2}} 
\nonumber \\
&&\times
\left\{
\frac{1}{(2\pi)^6}\frac{d\vec{k}_+}{2E_+}\frac{d\vec{k}_-}{2E_-} 
\frac{1}{q^4}f^{\mu\nu}(k_+,k_-)F^{q\bar q}_{\mu\nu}(p_q,p_{\bar{q}})
\right\} .
\label{eq:dy-crst-pn-1}
\end{eqnarray}
The quantity in the bracket $\{ ...\}$ of the above equation is
precisely what is in the bracket $\{ ...\}$ of Eq.~(\ref{eq:dy-qbarq-crst}) 
for the $q\bar{q}\to \mu^+\mu^-$ process. 
Accounting for the difference
in flux factors and extending Eq.~(\ref{eq:dy-crst-pn-1}) to
include the $q\leftrightarrow \bar{q}$ interchange term,
the full expression of the $pN$ DY process is
\begin{eqnarray}
\frac{d{\sigma}^{pN}(p_p,p_N)}{dq^2}
&=&
\sum_{q}\int d\vec{p}_q d\vec{p}_{\bar{q}}
[f^q_{\vec{p}_p} (\vec{p}_q) f^{\bar{q}}_{\vec{p}_N} (\vec{p}_{\bar{q}})
+ f^q_{\vec{p}_N} (\vec{p}_q) f^{\bar{q}}_{\vec{p}_p} (\vec{p}_{\bar{q}})] 
\nonumber \\
&& \times
\frac{[(p_q\cdot p_{\bar{q}})^2-m^4_q]^{1/2}}{[(p_p\cdot p_N)^2-m_p^2m_N^2)]^{1/2}}
\frac{E_pE_N}{E_qE_{\bar{q}}} \frac{d{\sigma}^{q\bar{q}}(p_q,p_{\bar{q}})}{dq^2} ,
\label{eq:dy-pN}
\end{eqnarray}
where {$d\sigma^{q\bar{q}}(p_q,p_{\bar{q}})/dq^2$} is the
$q\bar{q}$ DY cross section, as defined by  Eq.~(\ref{eq:dy-0}).

We now examine the physical meaning of the functions
$f^q_{\vec{p}_p} (\vec{p}_q)$ and $f^{\bar{q}}_{\vec{p}_N} (\vec{p}_{\bar{q}})$ in Eq.~(\ref{eq:dy-pN}).
By using the definitions~(\ref{eq:pn-2}) for 
$\phi_{\vec{p}_p}(\vec{p}_{\bar{q}} ,\vec{p}_{X_p})$ and Eq.~(\ref{eq:q-dis}) for
$f^q_{\vec p_p} (\vec p_q)$, it is straightforward to show that
\begin{eqnarray}
f^q_{\vec{p}_p}(\vec{p}_q) 
&=& 
\frac{ \bra{\vec{p}_p} b^\dagger_{\vec{p}_q}b_{\vec{p}_q} \ket{\vec{p}_p}} 
     {\inp{\vec{p}_p}{\vec{p}_p}}\,.
\label{eq:probab}
\end{eqnarray}
Thus $f^q_{\vec{p}_p}(\vec{p}_q)$ is just the probability of finding a quark $q$ with momentum $\vec{p}_q$ in a
nucleon state $\ket{\vec{p}_p}$.
Note that this simple interpretation of $f^{q}_{\vec{p}_p}(\vec{p}_q)$ is
due to the use of the approximations Eqs.~(\ref{eq:delta-f}) and~(\ref{eq:app-2}).
If we depart from these two simplifications, we then need the spectral function
of the nucleon in terms of parton degrees of freedom to make
calculation for DY cross sections. Such information is not available
at the present time.

\begin{figure}[t]
\includegraphics[clip,width=0.5\textwidth]{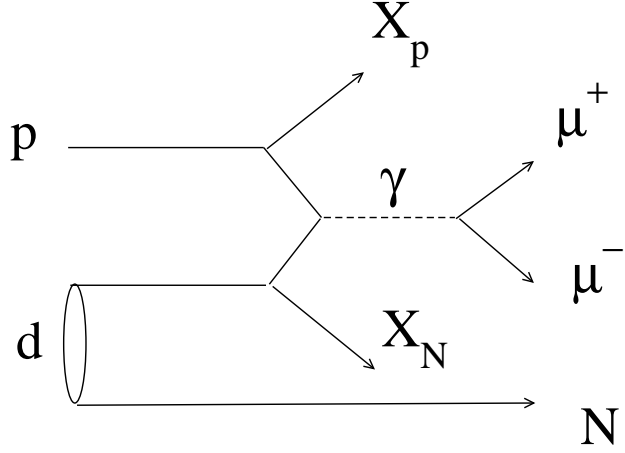}
\caption{
Impulse approximation of $pd$ DY process.
}
\label{fig:dy-mech-imp}
\end{figure}

\section{Proton-Deuteron DY cross section}
\label{sec:pdcs}

In this section, we derive formula to express the proton-deuteron ($pd$) 
DY cross sections in terms of $d\sigma^{q\bar{q}}(p_q,p_{\bar{q}})/dq^2$ of Eq.~(\ref{eq:dy-0}) 
for the elementary $q+\bar{q} \rightarrow \mu^+ +\mu^- $ process.
To simplify the presentation, we only explain the derivation of the formula 
for a quark $q$ emitted from the projectile $p$ and an anti-quark $\bar{q}$ from
the target $d$.
The terms from the interchange $q\leftrightarrow \bar{q}$ will be included only
in the final expression of the cross section.

For the $p(p_p) + d(p_d) \to \mu^+(k_+) + \mu^-(k_-) +X_p + X_d$ 
DY precess, Eq.~(\ref{eq:dy-crst}) gives
\begin{eqnarray}
d\sigma^{pd} 
&=&
\frac{(2\pi)^4}{4[(p_p\cdot p_d)^2 -m^2_pm^2_d]^{1/2}}
\frac{1}{(2\pi)^6}\frac{d\vec{k}_+}{2E_+}\frac{d\vec{k}_-}{2E_-}
\frac{1}{q^4}f^{\mu\nu}(k_+,k_-)F^{pd}_{\mu\nu}(p_p,p_d,q) .
\label{eq:dy-crst-pd}
\end{eqnarray}
To proceed, we need to define a model for generating the deuteron wave function.
Here we follow the $\pi NN$ studies~\cite{garcilazo,lee-nn}
to consider a nuclear model within which the deuteron wave function has two components
\begin{eqnarray}
\ket{\Psi_d} &=& \ket{\Phi_d} + \ket{\phi_{\pi NN}}  ,
\label{eq:d-2cp}
\end{eqnarray}
where $\Phi_d$ is the $NN$ component.
In the following two subsections, we derive formulas
for calculating the contribution from each component of the
deuteron wave function $\Psi_d$ to the $pd$ DY cross sections.

\subsection{Contribution from nucleons}
\label{sec:dycs-1}

We assume that the $pd$ DY process takes place on each of the nucleons 
in the deuteron, as illustrated in Fig.~\ref{fig:dy-mech-imp}.
In this impulse approximation, the hadronic tensor for a deuteron target 
can be obtained by simply extending Eq.~(\ref{eq:pN-tensor}) for the $pN$ to 
include a spectator nucleon state $\ket{p_s}$ in the sum over the final hadronic states. 
We thus have
\begin{eqnarray}
F^{pd}_{\mu\nu}(p_p,p_d,q)
&=&
(2\pi)^6 (2E_p)(2E_d) \sum_{X_p,X_N} \int d\vec{p}_s
d\vec{p}_{X_p}  d\vec{p}_{X_N}
\delta^4(p_p+p_d-p_{X_p}-p_{X_N}-p_s-q)
\nonumber \\
&&\times 
\bra{\Phi_{p_d}\vec p_p } J_\mu(0) \ket{\vec p_{X_p} \vec p_{X_N} \vec p_s}
\bra{\vec p_s \vec p_{X_N} \vec p_{X_p} } J_\nu(0) \ket{\vec p_p\Phi_{p_d}} \,,
\label{eq:pd-tensor-0}
\end{eqnarray}
where $\ket{\Phi_{p_d}}$ is the $NN$ component of a deuteron moving with a momentum $p_d$. 
We expand $\ket{\Phi_{p_d}}$ in terms of $NN$ plane-wave states
\begin{eqnarray}
\ket{\Phi_{p_d}} &=& 
\int d\vec{p}_N d\vec{p}_2 \Phi_{p_d}(\vec{p}_N)
\delta(\vec{p}_d-\vec{p}_N-\vec{p}_2)
\ket{\vec{p}_N\vec{p}_2} .
\label{eq:d-wf}
\end{eqnarray}
Keeping only the contributions due to a parton in $\ket{\vec{p}_N}$ of 
the above expansion of ${\Phi_{p_d}}$ and a parton from projectile state $\ket{\vec{p}_p}$, 
the current matrix element in  Eq.~(\ref{eq:pd-tensor-0}) becomes
\begin{eqnarray}
\bra{\Phi_{p_d}\vec p_p } J_\mu(0) \ket{\vec p_{X_p} \vec p_{X_N} \vec p_s}  =
 \int d \vec{p}_N \Phi^*_{p_d}(\vec{p}_N)
\delta(\vec{p}_d-\vec{p}_N-\vec{p}_s) 
\bra{\vec p_N \vec p_p} J_\mu(0) \ket{\vec p_{X_p}\vec p_{X_N}} .
\label{eq:c-mxd}
\end{eqnarray}
By using Eq.~(\ref{eq:c-mxd}), Eq.~(\ref{eq:pd-tensor-0}) can then be written as
\begin{eqnarray}
F^{pd}_{\mu\nu}(p_p,p_d,q)
&=&
\int d\vec{p}_N |\Phi_{p_d}(p_N)|^2\frac{(2E_p)(2E_d)}{(2E_p)(2E_N)}
\nonumber\\
&&\times
\{ (2\pi)^6 (2E_p)(2E_N) 
\nonumber \\
&& \times 
\sum_{X_p,X_N}\int d\vec{p}_{X_p} d\vec{p}_{X_N}
\delta^4(p_p+p_N-p_{X_p}-p_{X_N}-q) 
\nonumber \\
&&\times
\bra{\vec p_N\vec p_p } J_\mu(0) \ket{\vec p_{X_p} \vec p_{X_N}}
\bra{\vec p_{X_N} \vec p_{X_p}} J_\nu(0) \ket{\vec p_p\vec p_N} \}  .
\label{eq:pd-tensor}
\end{eqnarray}
We see that the quantity within the  bracket $\{...\}$ in the above equation 
is identical to $F^{pN}_{\mu\nu}(p_p,p_N,q)$ of Eq.~(\ref{eq:pN-tensor}).
We then have 
\begin{eqnarray}
F^{pd}_{\mu\nu}(p_p,p_d,q)
&=&
\int d\vec{p}_N \rho_{p_d}(\vec{p}_N)
\frac{E_p E_d}{E_pE_N} F^{pN}_{\mu\nu}(p_p,p_N,q) ,
\label{eq:d-tensor-a}
\end{eqnarray}
where
\begin{eqnarray}
\rho_{p_d}(\vec{p}_N) = |\Phi_{p_d}(\vec{p}_N)|^2 .
\end{eqnarray}
By using Eq.~(\ref{eq:d-wf}), one can show that
\begin{eqnarray}
\rho_{p_d}(\vec{p}_N) 
&=&
\bra{\Phi_{p_d}} b^\dagger_{{\vec{p}_N}}b_{{\vec{p}_N}} \ket{\Phi_{p_d}} , 
\label{eq:rho-n}
\end{eqnarray}
where $ b^\dagger_{{\vec{p}_N}}$ is the creation operator for a nucleon with momentum $\vec{p}_N$.
Thus $\rho_{p_d}(\vec{p}_N)$ is the nucleon momentum distribution
in a $moving$ deuteron with momentum ${p}_d$.
We will present formula for calculating ${\rho}_{p_d}(\vec{p}_N)$ in Sec.~\ref{sec:dist}.

By using Eq.~(\ref{eq:d-tensor-a}), Eq.~(\ref{eq:dy-crst-pd}) becomes
\begin{eqnarray}
d\sigma^{pd} 
&=&
\frac{(2\pi)^4}{4[(p_p\cdot p_d)^2 -m^2_pm^2_d]^{1/2}}
\sum_{N=p,n}\int  d\vec{p}_N \rho_{p_d}(\vec{p}_N) \frac{E_p E_d}{E_pE_N}
\nonumber \\
&&\times
\left\{
\frac{1}{(2\pi)^6}\frac{d\vec{k}_+}{2E_+}\frac{d\vec{k}_-}{2E_-}
\frac{1}{q^4}f^{\mu\nu}(k_+,k_-) F^{pN}_{\mu\nu}(p_p,p_N,q) 
\right\} .
\label{eq:dy-crst-pd-1}
\end{eqnarray}
The quantity within the bracket $\{...\}$  of the above equation is
exactly what is in the bracket $\{...\}$ of
Eq.~(\ref{eq:dy-crst-pn}). Accounting for the difference in flux factor,
we obviously can write
\begin{eqnarray}
\frac{d{\sigma}^{pd}(p_p,p_d)}{dq^2}
&=&\sum_{N=p,n}\int  d\vec{p}_N \rho_{p_d}(\vec{p}_N)
\frac{[(p_p\cdot p_N)^2-m^2_pm^2_N]^{1/2}}
{[(p_p\cdot p_d)^2-m_p^2m_d^2]^{1/2}}
\frac{E_p E_d}{E_pE_N} \frac{d{\sigma}^{pN}(p_p,p_N)}{dq^2}  ,
\label{eq:dy-pd-0}
\end{eqnarray}
where $d\sigma^{pN}(p_p,p_N)/dq^2$ are given in Eq.~(\ref{eq:dy-pN}).

Substituting Eq.~(\ref{eq:dy-pN}) into Eq.~(\ref{eq:dy-pd-0}), we have
\begin{eqnarray}
\frac{d{\sigma}^{pd}(p_p,p_d)}{dq^2}
&=&
\sum_{N=p,n}\int  d\vec{p}_N \rho_{p_d}(\vec{p}_N) 
\sum_{q}\int d\vec{p}_q d\vec{p}_{\bar{q}}
\frac{[(p_q\cdot p_{\bar{q}})^2-m^4_q]^{1/2}}
{[(p_p\cdot p_d)^2-m_p^2m_d^2]^{1/2}}
\frac{E_p E_d}{E_qE_{\bar{q}}} 
\nonumber \\
&& \times
[f^q_{\vec p_p} (\vec p_q) f^{\bar{q}}_{\vec p_N} (\vec p_{\bar{q}})+
f^q_{\vec p_N} (\vec p_q) f^{\bar{q}}_{\vec p_p} (\vec p_{\bar{q}}) ] 
\frac{d{\sigma}^{q\bar{q}}(p_q,p_{\bar{q}})}{dq^2}  \,.
\label{eq:dy-pd}
\end{eqnarray}

\subsection{Contribution from pions}
\label{sec:dycs-2}

In the impulse approximation, the contribution from the 
$\pi NN$ component of Eq.~(\ref{eq:d-2cp}) to the $pd$ DY cross sections can be
derived by using the similar procedures detailed in  the previous subsection.
We find  that the resulting formula can be obtained from Eq.~(\ref{eq:dy-pd}) by
changing the momentum distribution and parton distributions
for the nucleon to those for the pion. 
Explicitly, we have
\begin{eqnarray}
\frac{d{\sigma}^{pd}_\pi (p_p,p_d)}{dq^2}
&=&
\int  d\vec{k}_\pi \rho_{p_d}(\vec{k}_\pi)
\sum_{q}\int d\vec{p}_q d\vec{p}_{\bar{q}}
\frac{[(p_q\cdot p_{\bar{q}})^2-m^4_q]^{1/2}}{[(p_p\cdot p_d)^2-m_p^2m_d^2)]^{1/2}}
\frac{E_p E_d}{E_qE_{\bar{q}}} 
\nonumber \\
&& \times
[f^q_{\vec p_p} (\vec p_q) f^{\bar{q}}_{\vec k_\pi} (\vec p_{\bar{q}})+
f^q_{\vec k_\pi} (\vec p_q) f^{\bar{q}}_{\vec p_p} (\vec p_{\bar{q}}) ]
\frac{d{\sigma}^{q\bar{q}}(p_q,p_{\bar{q}})}{dq^2} \,,
\label{eq:dy-pd-pi}
\end{eqnarray}
where $f^q_{\vec k_\pi}$ and $f^{\bar{q}}_{\vec k_\pi}$
are  the PDFs for the
pion, and the pion momentum distribution in a moving deuteron with momentum $p_d$ is defined by
\begin{eqnarray}
\rho_{p_d}(\vec{k}_\pi) 
&=& 
\bra{\phi_{\pi NN. p_d}} a^\dagger_{\vec{k}_\pi}a_{\vec{k}_\pi} 
\ket{\phi_{\pi NN, p_d}}.
\label{eq:rho-pi}
\end{eqnarray}
where $ a^\dagger_{\vec{k}_\pi}$ is the creation operator for a
pion with momentum $\vec{k}_\pi$.
The calculation of $\rho_{p_d}(\vec{k}_\pi)$ from a $\pi NN$ model will be
 given in
the next section.

\section{Calculations of nucleon and pion momentum distributions}
\label{sec:dist}

We first describe a nuclear model within which the procedure for calculating the nucleon and pion momentum
distributions in the rest frame of the deuteron is described.
We then explain how the these distributions can be used to
calculate the momentum distributions in a fast moving deuteron, which
are the input to our calculations of Eqs.~(\ref{eq:dy-pd}) and~(\ref{eq:dy-pd-pi}).

\subsection{$\pi NN$ Model}
\label{sec:dist-1}

We follow the $\pi NN$ studies~\cite{garcilazo,lee-nn} 
to consider a nuclear model defined by the following  Hamiltonian 
\begin{eqnarray}
H=H_0 + V_{NN} +\, [\, H_{\pi NN}+H^\dagger_{\pi NN}\,]\,\,,
\label{eq:tot-h}
\end{eqnarray}
where $H_0$ is the sum of free energy operators for $N$ and $\pi$, $V_{NN}$ is
a nucleon-nucleon  potential, and $H_{\pi NN}$ defines
the virtual  $N \to \pi N$ transition
\begin{eqnarray}
H_{\pi NN} &=& \sum_{i=1,2} h_{\pi NN}(i) ,
\label{eq:pinn-vert}
\end{eqnarray}
where $i$ denotes the $i$-th nucleon.
Starting with the standard pseudo-vector coupling, the vertex interaction
takes the following familiar form (omitting the spin-isospin indices)
\begin{eqnarray}
h_{\pi NN}(i) &=& 
\int d\vec{k} d\vec{p}_i d\vec{p}^{\,\,\prime}_i 
\,\, \delta(\vec{p}_i+\vec{k}-\vec{p}^{\,\,\prime}_i)
\,\,[ \,\, \ket{{\vec{p}_i} \vec{k}} F_{\pi NN}(\vec{p}_i,\vec{k})
 \bra{\vec{p}^{\,\,\prime}_i}\,\,] ,
\label{eq:v-h}
\end{eqnarray}
where $\ket{\vec{p}_i}$ and $\ket{\vec{k}}$ are the plane wave states
of the $i$-th nucleon and pion, respectively, and
\begin{eqnarray}
F_{\pi NN}(\vec{p},\vec{k})
&=&
-\frac{i}{(2\pi)^{3/2}}\frac{f_{\pi NN}}{m_\pi}
 \frac{1}{\sqrt{2E_\pi(\vec{k})}}\sqrt{\frac{m_N}{E_N(\vec{p})}}
\sqrt{\frac{m_N}{E_N(\vec{p}+\vec{k})}}
\bar{u}_{\vec{p}}\slash{k}\gamma_5 u_{\vec{p}+\vec{k}}F(\Lambda_{\pi NN},\vec{k}) ,
\nonumber\\
\label{eq:v-f}
\end{eqnarray}
Here $F(\Lambda_{\pi NN},k)$ is a form factor that satisfies
$F(\Lambda_{\pi NN},k) = 1$ at $k = i\kappa$ with 
$\kappa = m_\pi\sqrt{1-m_\pi^2/(4m_N^2)}$ being the pion
momentum at the nucleon pole of the $\pi N$ amplitude, and
its cutoff parameter $\Lambda_{\pi NN}$ 
can be determined in the fit to the $\pi N$ scattering data.

It follows from Eqs.~(\ref{eq:tot-h})-(\ref{eq:v-h})
that the  bound state $\ket{{\Psi}_d}$ in the deuteron rest 
frame ($\vec{p}_d =0$)
is defined by
\begin{eqnarray}
H\ket{\Psi_d} &=& E_d\ket{\Psi_d} ,
\label{eq:d-bs}
\end{eqnarray}
where the deuteron  wave function is normalized to $\inp{\Psi_d}{\Psi_d}=1$
and has two components
\begin{eqnarray}
\ket{{\Psi}_d}&=& \frac{1}{Z^{1/2}}[\ket{\phi_{NN}} + a \ket{\phi_{\pi NN}} ] \,.
\label{eq:d-wf-pinn}
\end{eqnarray}
Here $Z$ is a normalization factor and each component of the wave function
is normalized to 1:
$\inp{\phi_{NN}}{\phi_{NN}}=1$ and $\inp{\phi_{\pi NN}}{\phi_{\pi NN}}=1$.
By using the orthogonality condition 
$\inp{\phi_{NN}}{\phi_{\pi NN}}=0$, 
Eqs.~(\ref{eq:d-bs}) and~(\ref{eq:d-wf-pinn}) lead to
\begin{eqnarray}
a\bra{\phi_{\pi NN}}(H_0+V_{NN}) \ket{\phi_{\pi NN}} 
+ \bra{\phi_{\pi NN}} H_{\pi NN} \ket{\phi_{NN}} = aE_d ,
\label{eq:proj-pinn} 
\\
\bra{\phi_{ NN}} (H_0+V_{NN}) \ket{\phi_{ NN}} 
+ a\bra{\phi_{NN}} H^\dagger_{\pi NN} \ket{\phi_{\pi NN}}=E_d .
\label{eq:proj-nn}
\end{eqnarray}
From Eq.~(\ref{eq:proj-pinn}), we have
\begin{eqnarray}
a 
&=&
\frac{ \bra{\phi_{\pi NN}} H_{\pi NN} \ket{\phi_{NN}}}
{E_d- \bra{\phi_{ \pi NN}} (H_0+V_{NN}) \ket{\phi_{ \pi NN}}}.
\label{eq:pinn-cof}
\end{eqnarray}

It is a difficult three-body problem to solve Eqs.~(\ref{eq:proj-pinn})
and~(\ref{eq:proj-nn}) exactly and find a model of $V_{NN}$ to fit
the $NN$ scattering data and the deuteron bound state properties. 
Here we are simply guided by the results from the 
previous $\pi NN$ studies~\cite{garcilazo,lee-nn}. 
It was found that in the low energy region, 
the $\pi NN$ component is much weaker than the $NN$ component, 
and it is a good approximation to neglect the matrix element of 
$\bra{\phi_{\pi NN}} V_{NN} \ket{\phi_{\pi NN}}$ in Eq.~(\ref{eq:pinn-cof}).
We then have from Eq.~(\ref{eq:proj-nn}),
\begin{eqnarray}
&& 
\bra{\phi_{NN}} (H_0+V_{NN}) \ket{\phi_{NN}} +
\frac{\bra{\phi_{NN}} H^\dagger_{\pi NN} \ket{\phi_{\pi NN}}
\bra{\phi_{\pi NN}} H_{\pi NN} \ket{\phi_{NN}}}
{E_d-\bra{\phi_{\pi NN}} H_0 \ket{\phi_{\pi NN}}} = E_d ,
\label{eq:eff-nn}
\end{eqnarray}
and the coefficient $a$ of the total wave function~(\ref{eq:d-wf-pinn}) is 
\begin{eqnarray}
a=
\frac{ \bra{\phi_{\pi NN}} H_{\pi NN} \ket{\phi_{NN}}}
{E_d - \bra{\phi_{ \pi NN}} H_0 \ket{\phi_{ \pi NN}}}.
\label{eq:wf-coef}
\end{eqnarray}
If we further assume that the pion loop (pion is emitted and absorbed by the same nucleon)
in the second term of Eq.~(\ref{eq:eff-nn}) can be absorbed in the physical nucleon mass,
Eq.~(\ref{eq:eff-nn}) is equivalent to the following Schr\"odinger equation 
\begin{eqnarray}
[\, H_0 + V_{NN}+V^{\text{opep}}_{NN}(E_d)\,]\,\ket{\phi_{NN}} = E_d \ket{\phi_{NN}},
\label{eq:eff-nn-wf}
\end{eqnarray}
with the one-pion-exchange potential defined by
\begin{eqnarray}
V^{\text{opep}}_{NN}(E_d)
&=&
\sum_{i\neq j}h^\dagger_{\pi NN}(i) 
\frac{\ket{\phi_{\pi NN}}\bra{\phi_{\pi NN}}}{E_d - {\bra{\phi_{\pi NN}} H_0 \ket{\phi_{\pi NN}}}}
h_{\pi NN}(j) .
\label{eq:opep}
\end{eqnarray}

By using Eq.~(\ref{eq:pinn-vert}) for $H_{\pi NN}$, Eq.~(\ref{eq:wf-coef}) leads to
\begin{eqnarray}
|a|^2 &=&\bra{\phi_{NN}}[ \rho^{\text{loop}}_\pi(E_d) + \rho^{\text{exc}}_\pi(E_d) ]\ket{\phi_{NN}} ,
\label{eq:a2}
\end{eqnarray}
where
\begin{eqnarray}
\rho^{\text{exc}}_\pi(E_d) 
&=& 
\sum_{i\neq j}h^\dagger_{\pi NN}(i)
\frac{\ket{\phi_{\pi NN}}\bra{\phi_{\pi NN}}} 
{(E_d - {\bra{\phi_{\pi NN}} H_0 \ket{\phi_{\pi NN}}})^2} 
h_{\pi NN}(j),
\label{eq:rho-exc}
\\
\rho^{\text{loop}}_\pi(E_d) 
&=& 
\sum_{i}h^\dagger_{\pi NN}(i)
\frac{\ket{\phi_{\pi NN}}\bra{\phi_{\pi NN}}}{(E_d - {\bra{\phi_{\pi NN}} H_0 \ket{\phi_{\pi NN}}})^2} 
h_{\pi NN}(i).
\end{eqnarray}
From Eqs.~(\ref{eq:opep}) and~(\ref{eq:rho-exc}), we then have the following interesting relation
\begin{eqnarray}
\rho^{\text{exc}}_\pi (E_d)
&=& 
-\frac{d}{dE_d} V^{\text{opep}}_{NN}(E_d) .
\label{eq:rho-opep-rel}
\end{eqnarray}
Note that the relation Eq.~(\ref{eq:rho-opep-rel}) is identical to
that used in Ref.~\cite{fpw} to calculate the so-called pion-excess, 
except that a non-relativistic form of $V^{\text{opep}}_{NN}$ is used in their calculations.

To see the physical meaning of $\rho^{\text{loop}}_\pi(E_d)$ and
$\rho^{\text{exc}}_\pi(E_d)$, we define the
pion number $N_\pi$ in the deuteron rest frame as
\begin{eqnarray}
N_\pi &=& \int d\vec{k} \rho_\pi(\vec{k}) ,
\end{eqnarray}
with 
\begin{eqnarray}
\rho_\pi(\vec{k}) &=& \bra{ \Psi_d } a^\dagger_{\vec{k}} a_{\vec{k}} \ket{ \Psi_d } ,
\end{eqnarray} 
From Eq.~(\ref{eq:d-wf-pinn}), we then get
\begin{eqnarray}
N_\pi &=& \frac{1}{Z} |a|^2 .
\label{eq:num-pi}
\end{eqnarray}
By using Eqs.~(\ref{eq:a2}) and~(\ref{eq:num-pi}), we can define
\begin{eqnarray}
N_\pi &=& N^{0}_\pi + N^{\text{exc}}_\pi,
\end{eqnarray}
where
\begin{eqnarray}
N^{0}_\pi &=&
\frac{1}{Z} \bra{\phi_{NN}} \rho^{\text{loop}}_\pi(E_d) \ket{\phi_{NN}} 
\nonumber \\
&=&
\int d\vec{k} \rho^{0}_\pi(\vec{k})\label{eq:pi-dir-00}, 
\\
&& 
\nonumber \\
N^{\text{exc}}_\pi &=&
\frac{1}{Z} \bra{\phi_{NN}} \rho^{\text{exc}}_\pi(E_d) \ket{\phi_{NN}} 
\nonumber \\
&=&
\int d\vec{k} \rho^{\text{exc}}_\pi(\vec{k}).
\label{eq:pi-exc-00}
\end{eqnarray}
In the DY and DIS calculations, the contributions from $\rho^{0}_\pi(\vec{k})$
are included in the meson cloud contributions to the nucleon parton 
distributions. Only $\rho^{\text{exc}}_\pi(\vec{k})$ is needed in our
calculation of pion contribution to proton-deuteron $DY$ process. 
This assumption is  similar to that used in
the calculation of pion-excess contribution~\cite{fpw}
to DIS cross sections~\cite{bc}.

\begin{figure}[t]
\includegraphics[clip,width=0.5\textwidth]{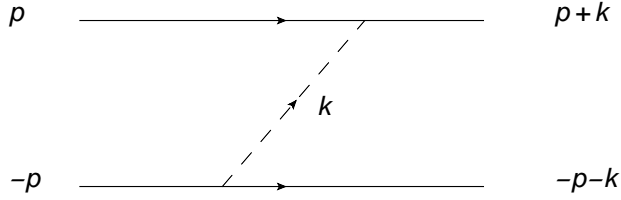}
\caption{One-pion-exchange interaction in the center of mass frame of $NN$.}
\label{fig:opep}
\end{figure}

To calculate $\rho^{\text{exc}}_\pi(\vec{k})$, we use Eq.~(\ref{eq:rho-opep-rel}) by 
first calculating the matrix element of one-pion-exchange potential~(\ref{eq:opep}) 
in the the rest frame of the deuteron.
From the kinematics shown in Fig.~\ref{fig:opep}, we have
\begin{eqnarray}
\bra{\phi_{NN}} V^{\text{opep}}_{NN}(E_d) \ket{\phi_{NN}} 
&=&
\int d\vec{k} \bra{\phi_{NN}} V^{\text{opep}}_{NN} (\vec{k}, E_d) \ket{\phi_{NN}}
\label{eq:opep-mx}
\end{eqnarray}
where the contribution from the pion with momentum $\vec{k}$ is
\begin{eqnarray}
\bra{\phi_{NN}} V^{\text{opep}}_{NN}(\vec{k}, E_d) \ket{\phi_{NN}}
&=& 
\int d\vec{p} \phi^*_{NN}(\vec{p}+\vec{k}) 
F^*_{\pi NN}(\vec{p},\vec{k})
\frac{1}{E_d-E_N(\vec{p})-E_N({\vec{p}+\vec{k}})-E_\pi(\vec{k})} 
\nonumber \\
&&\times
F_{\pi NN}({-\vec{p}-\vec{k}},\vec{k})  \phi_{NN}(\vec{p}) .
\nonumber
\end{eqnarray}
By using Eqs.~(\ref{eq:rho-opep-rel}) and (\ref{eq:v-f}) and 
including spin-isospin indices, we readily get
\begin{eqnarray}
\rho_\pi^{\text{exc}}(\vec{k}) 
&=& 
\frac{1}{Z}\,[\,-\frac{d}{dE_d} \bra{\phi_{NN}} V^{\text{opep}}_{NN}( \vec{k}, E_d) \ket{\phi_{NN}}\,]
\nonumber \\ 
&=& 
\frac{2}{Z}\int d\vec{p}\frac{1}{(2\pi)^3} \left(\frac{f_{\pi NN}}{m_\pi}\right)^2
\frac{1}{2E_\pi(\vec k)} \frac{m_N}{E_N(\vec{p})} \frac{m_N}{E_N(\vec{p}+\vec{k})} 
\left[\frac{1}{E_d- E_N(\vec{p})-E_N(\vec{p}+\vec{k})-E_\pi (\vec{k})}\right]^2 
\nonumber \\
&&\times 
\sum_{m_{s_1}m_{s_2} m_{s'_1} m_{s'_2}}
\phi^{J_dM_{J_d}*}_{NN}(\vec{p}+\vec{k},m_{s_1},m_{s_2})
\bar{u}_{\vec{p}+\vec{k},m_{s_1}} \slash{k}\gamma_5 u_{\vec{p},m_{s'_1}}
\bar{u}_{-\vec{p}-\vec{k},m_{s_2}}\slash{k}\gamma_5u_{-\vec{p}, m_{s'_2}}
\nonumber \\
&& \times 
[F(\Lambda_{\pi NN},\vec{k})]^2
\phi^{J_dM_{J_d}}_{NN}(\vec{p},m_{s'_1},m_{s'_2})
\bra{T_d M_d} I(\tau_1, \tau_2 )\ket{T_d M_d},
\label{eq:pi-dis}
\end{eqnarray}
where the overall factor 2 comes from summing up two possible pion-exchange diagrams.
The isospin matrix element is
\begin{eqnarray}
\bra{T_d M_d} I(\tau_1, \tau_2 )\ket{T_d M_d}
&=& \sum_{\alpha= -1,0,+1}\sum_{T' M'_T}
\bra{T_d M_d}  \tau_\alpha(1) 
\ket{[\pi_\alpha N_1 N_2]_{T' M'_T}}
\nonumber\\
&&
\qquad \qquad
\times
\bra{[\pi_\alpha N_1 N_2]_{T' M'_T}} 
\tau_{-\alpha}(2)\ket{T_d M_d}
(-1)^\alpha,  \nonumber
\end{eqnarray}
with
\begin{eqnarray}
\ket{T M_T}
&=& \sum_{m_1,m_2} \inp{\frac{1}{2}\frac{1}{2} m_1 m_2}{T M_T} \ket{m_1,m_2} ,
\\
\ket{[\pi_\alpha N_1 N_2]_{T' M'_T}} 
&=& \sum_{t}\ket{[\pi_\alpha [N_1N_2]_t]_{T' M'_T}} \nonumber \\
&=& \sum_{t} \sum_{ m_1,m_2}
\inp{\frac{1}{2}\frac{1}{2} m_1 m_2}{t m_t}\inp{t 1 m_t m_\alpha}{T'M'_T}
\ket{m_1,m_2} ,
\label{eq:ivector}
\end{eqnarray}
where $\inp{j_1j_2m_1m_2}{jm}$ is the Clebsch-Gordon coefficient.

For deuteron $T=M_T=0$ and only $T' = M'_T=0$ for $\pi NN$ should be kept, 
we then get $\bra{T_d M_d} I(\tau_1, \tau_2 )\ket{T_d M_d}=-1/3$.
The spin-orbital part of the
deuteron wave function in Eq.~(\ref{eq:pi-dis}) can be expanded as
\begin{eqnarray}
\phi^{J_dM_{J_d}}_{NN}(\vec{p}, m_{s_1}, m_{s_2}) &=&
\sum_{LM_L} \inp{LSM_L m_s}{J_dM_{{J_d}}}
\inp{\frac{1}{2}\frac{1}{2} m_{s_1} m_{s_2}}{Sm_S}
u_L(p) Y_{LM_L}(\hat{p})  ,
\label{eq:d-pw}
\end{eqnarray}
where the radial wave functions are normalized as
\begin{eqnarray}
\int_0^\infty p^2dp [u^2_0(p)+u^2_2(p)] &=& 1 .
\end{eqnarray}
Because the $\pi NN$ component is much smaller, the normalization factor is $Z \sim 1$. 
We will use $Z=1$ for calculating $\rho^{\text{exc}}_\pi(\vec{k})$ of Eq.~(\ref{eq:pi-dis}).
In the same approximation, we                                          
will use the deuteron radial wave functions
$u_0(p)$ and $u_2(p)$ generated from the available
realistic $NN$ potentials such as ANL-V18~\cite{v14}.

Neglecting the small contribution from $\pi NN$ component,
the nucleon momentum distribution in the rest frame of the deuteron can be written as
\begin{eqnarray}
\rho_N(\vec{p})
&=& 
\bra{\Phi_d} b^\dagger_{\vec{p}} b_{\vec{p}} \ket{\Phi_d} 
\nonumber \\
&\sim& 
\frac{1}{Z} \bra{\phi_{NN}} b^\dagger_{\vec{p}} b_{\vec{p}} \ket{\phi_{NN}} .
\label{eq:n-dis}
\end{eqnarray}
By using Eq.~(\ref{eq:d-pw}) and setting $Z \sim 1$, we obtain
\begin{eqnarray}
\rho_N(\vec{p}) \sim \frac{1}{4\pi}[u^2_0(p)+u^2_2(p)] .
\label{eq:deut-rhop}
\end{eqnarray}

\subsection{Momentum distributions in a moving deuteron}
\label{sec:dist-2}

In the calculation of proton-deuteron DY cross sections,
the momentum distributions $\rho_{p_d}(\vec{p}_N)$
in Eq.~(\ref{eq:dy-pd}) and $\rho_{p_d}(\vec{k}_\pi)$ in Eq.~(\ref{eq:dy-pd-pi}) 
are defined in a fast moving deuteron with a momentum $p_d$.
To calculate such momentum distributions, we first note that
the particle number in a system is independent of the frame.
We thus  have the following frame 
independent normalization condition
\begin{eqnarray}
N_a
= \int d \vec p^{\,\,'} \rho_{p_d}(\vec p^{\,\,'})
= \int d\vec p  \rho_{p^\circ_d}(\vec{p}) ,
\label{eq:ltran-1}
\end{eqnarray}
where $N_a$ is the number of the considered particle $a =N$~or~$\pi$ in the deuteron, and
the deuteron momenta (set $\vec{p}_d$ in the z-direction)
in the moving frame and the rest frame are, respectively,
\begin{eqnarray}
p_d &=& (E_d(\vec p_d), 0, 0, p_{d}^z ), \\
p^\circ_d &=& (m_d, 0,0,0) .
\end{eqnarray}
The nucleon momenta in Eq.~(\ref{eq:ltran-1}) are 
related by the Lorentz transformation defined by the
velocity  $\beta =  P_{z}/E_d(\vec{P}_d)$ of the moving frame.
Explicitly, we have
\begin{eqnarray}
p_z
&=&
\frac{E_d(\vec p_d )}{m_d}
\left[ p'_z - \frac{p_{d}^z}{E_d(\vec p_d )} E_N(\vec p') \right] ,
\label{eq:lt-pz} 
\\
E_N(p) 
&=&
\frac{E_d(\vec p_d )}{m_d} 
\left[ E_N(\vec p')- \frac{p_d^z}{E_d(\vec p_d)}p'_z \right] ,
\label{eq:lt-e}
\\
\vec{p}_\bot
&=& 
\vec{p}^{\,\,'}_\bot .
\label{eq:lt-pper}
\end{eqnarray}
The above equations lead to the following Lorentz invariant relation 
\begin{eqnarray}
\frac{d\vec{p}^{\,\,'}}{E_N(\vec p')} &=& \frac{d\vec{p}}{E_N(\vec p)} .
\label{eq:nord}
\end{eqnarray}
By using Eqs.~(\ref{eq:ltran-1}) and~(\ref{eq:nord}), we thus have
\begin{eqnarray}
\rho_{p_d}(\vec{p}^{\,\,'}) &=& \frac{ E_N(\vec p)}{E_N(\vec p')}\rho_{p^\circ_d}(\vec{p}) .
\label{eq:lt-dent}
\end{eqnarray}
With Eqs.~(\ref{eq:lt-pz})-(\ref{eq:lt-pper}), we can use
Eq.~(\ref{eq:lt-dent}) to get
$\rho_{p_d}(\vec{p}')$ from the momentum distribution 
$\rho_{p^\circ_d}(\vec{p})$ in the rest frame of the deuteron;
$\rho_{p^\circ_d}(\vec{p})$ can be calculated from 
the momentum distributions in the deuteron rest frame:
$\rho^{\text{exc}}_\pi(\vec{k})$  of Eq.~(\ref{eq:pi-dis}) for pions and
$\rho_N(\vec{p})$ of Eq.~(\ref{eq:deut-rhop}) for nucleons.

Here we mention that the relation~(\ref{eq:lt-dent}) for a two-nucleon
system can be explicitly
derived from the definition~(\ref{eq:rho-n}) within the
relativistic quantum mechanics developed by Dirac,
as reviewed in Ref.~\cite{polyzou}.

\section{Calculation Procedures}
\label{sec:numerical}

In this section, we develop procedures to apply the
formula presented in previous sections to investigate the
Fermi motion and pion-exchange effects on the ratio
$R_{pd/pp}=\sigma^{pd}/(2\sigma^{pp})$ between the $pd$ and
and $pp$ DY cross sections. 
Our first task is to relate our momentum variables $p_p$, $p_T$ and $q$
to the variables used in the analysis~\cite{peng2001} of the available data.
This will be given in Sec.~\ref{sec:numerical-1}. 
The procedures for calculating DY cross sections 
are given for $pp$ in Sec.~\ref{sec:numerical-2} and for $pd$ in Sec.~\ref{sec:numerical-3}.

\subsection{Kinematical variables for DY cross sections}
\label{sec:numerical-1}

It is common~\cite{peng2001} to use the collinear approximation to define the parton momentum:
\begin{eqnarray}
p_{q_p}= x_1 p_p \label{eq:x1x2-a} \,,\\
p_{q_T}=x_2 p_T  \label{eq:x1x2-b} \,, 
\end{eqnarray}
where $p_{q_p}$ ($ p_{q_T})$ is the momentum of a parton in the projectile (target), 
and $x_1$ and $x_2$ are scalar numbers.
The momentum $q$ of the virtual photon in 
the $q  \bar{q} \to \gamma \to \mu^+  \mu^-$, as seen in Fig.~\ref{fig:dy-mech}. is
\begin{eqnarray}
q&=&p_{q_p}+p_{q_T}=x_1p_p+x_2p_T .
\label{eq:q}
\end{eqnarray}
In the considered very  high energy region $E_p > 100 $ GeV,
the masses of projectile ($p^2_p=m^2_p$) and target ($p^2_T=m^2_T$)
can be neglected and hence $s=(p_p+p_T)^2 \sim  2p_p\cdot p_T$, 
$p_T\cdot q \sim x_1 p_p\cdot p_T$, and $ p_p\cdot q \sim x_2 p_p\cdot p_T$. 
We thus have the following relations
\begin{eqnarray}
x_1&\sim&\frac{2q\cdot p_T}{s},  \label{eq:x1-0}
\\
x_2&\sim&\frac{2q\cdot p_p}{s}.  \label{eq:x2-0}
\end{eqnarray}

It is most convenient to perform calculations in terms of $x_1$ and $x_2$
in the center of mass system in which the projectile is in the  $z$ direction
and the target in $-z$ direction:
\begin{eqnarray}
p_p &=& (\sqrt{p^2+m^2_p},0,0,p)\sim (p,0,0,p), 
\label{eq:pp-cm} \\
p_T &=& (\sqrt{p^2+m^2_T},0,0,-p) \sim (p,0,0,-p). 
\label{eq:pt-cm}
\end{eqnarray}
With the choices (\ref{eq:pp-cm}) and~(\ref{eq:pt-cm}), we have
\begin{eqnarray}
s&=& (p_p+p_T)^2 \sim 4p^2 , 
\nonumber \\
M^2 &=& q^2= (x_1p_p+x_2p_T)^2 \sim 4 x_1x_2 p^2 .
\nonumber
\end{eqnarray}
The above two equations lead the simple relation
\begin{eqnarray}
x_1x_2\sim \frac{M^2}{s} .
\label{eq:x1x2-ms}
\end{eqnarray}
By using Eqs.~(\ref{eq:x1-0})-(\ref{eq:pt-cm}), we can define a useful variable $x_F$ 
\begin{eqnarray}
x_F&=&x_1-x_2 \sim \frac{2(p_T-p_p)\cdot q}{s} 
\nonumber \\
&\sim&\frac{2\sqrt{s}\hat{z} \cdot \vec{q}}{s}
=\frac{2\sqrt{s}\hat{p}_p \cdot \vec{q}}{s} .
\end{eqnarray}
In the notation of Ref.~\cite{peng2001}, we write
\begin{eqnarray}
x_F\sim\frac{p^\gamma_{\parallel}}{\sqrt{s}/2} ,
\label{eq:xf}
\end{eqnarray}
where $p^\gamma_{\parallel}=\hat{p}_p \cdot \vec{q}$ is clearly the longitudinal 
component of the intermediate photon momentum with respect to the projectile 
in the center of mass frame.
Experimentally, $s$, $M$, $x_F$, and $d\sigma/(dM dx_F)$ are measured.
With the relation~(\ref{eq:x1x2-ms}), we certainly can determine
the corresponding $x_1$, $x_2$ and $d\sigma/(dx_1dx_2)$.
We thus will only  give the expression of $d\sigma/(dx_1dx_2)$ in the
following subsections.

\subsection{Calculation of $pp$ DY cross sections $d\sigma^{pp}/dx_1dx_2$}
\label{sec:numerical-2}

We now note that with the simplifications used in
defining the variable $x_1$ and $x_2$, as described above,
the flux factor associated with Eq.~(\ref{eq:dy-pN}) become 1.
Substituting Eq.~(\ref{eq:dy-0}) into Eq.~(\ref{eq:dy-pN}), the
DY cross section for $p(p_p)+N(p_N) \to \mu^+ + \mu^- +X_p +X_N $ with $N=p,n$
is then calculated from
\begin{eqnarray}
\frac{d\sigma^{pN}(p_p,p_N)}{dq^2}
&=&\sum_{q}\int d\vec{p}_q d\vec{p}_{\bar{q}}
[f^q_{p_p} (p_q) f^{\bar{q}}_{p_N} (p_{\bar{q}})+
f^q_{p_N} (p_q) f^{\bar{q}}_{p_p} (p_{\bar{q}}) ] \nonumber \\
&&\times
\frac{4\pi\alpha^2}{9q^2}\hat e^2_q\delta \left( q^2-(p_q+p_{\bar{q}})^2 \right) .
\label{eq:dy-pN-c0}
\end{eqnarray}
In the chosen center of mass frame, defined by Eqs.~(\ref{eq:pp-cm}) and~(\ref{eq:pt-cm}),
let us consider $\bar{q}$ in the target nucleon moving with
a momentum $p_N=(p^z_N,0,0,p^z_N)$.
In the precise collinear approximation, only $z$-component of the $\bar{q}$
momentum is defined by $p^z_N$. 
As defined by  Eq.~(\ref{eq:x2-0}),
we thus  write $\vec{p}_{\bar{q}}=(\vec{p}_{\bar{q}\perp}, p^z_{\bar{q}})$ where
\begin{eqnarray}
p^z_{\bar{q}} = x_2 p^z_N ,
\end{eqnarray}
and $\vec{p}_{\bar{q}\perp}$ can be arbitrary.
The integration over the $\bar{q}$ momentum distribution in the
target $N$ can then be written as
\begin{eqnarray}
\int d\vec{p}_{\bar{q}}f^{\bar{q}}_{\vec p_N}(\vec{p}_{\bar{q}})
&=& 
\int dx_2 f^{\bar{q}}_{N}(x_2) ,
\label{eq:pdf-1}
\end{eqnarray}
with
\begin{eqnarray}
f^{\bar{q}}_N(x_2) 
&=& 
p^z_N \int d\vec{p}_{\bar{q}\perp} f^{\bar q}_{\vec p_N}(x_2 p^z_N,\vec{p}_{\bar{q}\perp}) .
\label{eq:pdf-1a}
\end{eqnarray}
Similarly, we can define for the projectile proton
\begin{eqnarray}
\int d\vec{p}_{q}f^{q}_{\vec p_p}(\vec{p}_{q}) = \int dx_1 f^q_p(x_1) .
\label{eq:pdf-2}
\end{eqnarray}
By using Eqs.~(\ref{eq:pdf-1}) and~(\ref{eq:pdf-2}), 
Eq.~(\ref{eq:dy-pN-c0}) can be written as
\begin{eqnarray}
\frac{d\sigma^{pN}(p_p,p_N)}{dq^2}
&=&
\sum_{q} \int dx_1 dx_2
[f^q_{p} (x_1) f^{\bar{q}}_N (x_2) + f^{\bar{q}}_{p} (x_1) f^q_{N} (x_2) ]
\nonumber \\
&&\times
\frac{4\pi\alpha^2}
{9q^2}\hat e^2_q\delta\left(q^2-(p_q+p_{\bar{q}})^2\right) .
\label{eq:dy-pN-c}
\end{eqnarray}
Integrating the above equation over $dq^2$,  
we then obtain an expression of the cross section in terms
of $x_1$ and $x_2$ that are defined by experimental kinematics
\begin{eqnarray}
\frac{d\sigma^{pN}(p_p,p_N)}{dx_1dx_2}
&=&
\sum_{q} \frac{4\pi\alpha^2}{9(p_q+p_{\bar{q}})^2} \hat e^2_q
[f^q_{p} (x_1) f^{\bar{q}}_N (x_2) + f^{\bar{q}}_{p} (x_1) f^q_{N} (x_2) ] ,
\label{eq:dy-pp-c}
\end{eqnarray}
which is the same as Eq.~(\ref{eq:dy-exp}) used in the analysis of Ref.~\cite{peng2001}
since $(p_q+p_{\bar{q}})^2= q^2=M^2$ for the partonic process $q \bar{q}\to \gamma$.
Therefore we identify $f^{\bar{q}}_N(x)$, defined by Eq.~(\ref{eq:pdf-1a}),
with PDFs of the parton model [also for $f^{{q}}_N(x)$].  
To compare with the results of Ref.~\cite{peng2001},
we use PDFs of CETEQ5m~\cite{ceteq5m} in our calculations of Eq.~(\ref{eq:dy-pp-c}).

Equation~(\ref{eq:dy-pp-c}) for the $pp$ then obviously takes the following form
\begin{eqnarray}
\frac{d\sigma^{pp}(p_p,p_{N=p})}{dx_1dx_2}
&=&
\frac{4\pi\alpha^2}{9M^2}
\left[
\frac{4}{9}
\left(f^{u}_p(x_1)f^{\bar u}_{p}(x_2)+f^{\bar{u}}_p(x_1) f^{u}_{p}(x_2)\right) 
\right.
\nonumber \\
&& 
\left.
+\frac{1}{9}
\left(f^{d}_p(x_1)f^{\bar d}_{p}(x_2)+f^{\bar{d}}_p(x_1) f^{d}_{p}(x_2)\right) 
\right] .
\label{eq:dy-pp-cc}
\end{eqnarray}

\subsection{Calculation of $pd$ DY cross sections of $d\sigma^{pd}/(dx_1dx_2)$}
\label{sec:numerical-3}

We first consider the contributions from the nucleon momentum distribution $\rho_{p_d}(\vec{p}_N)$
to Eq.~(\ref{eq:dy-pd}) for the proton-deuteron DY cross sections .
With the simplifications used in
defining the variable $x_1$ and $x_2$, as described in the subsection V.A,
the flux factor associated with Eqs.~(\ref{eq:dy-pd}) become 1.
We thus only need to consider
\begin{eqnarray}
\frac{d\sigma^{pd}(p_p,p_d)}{dq^2}
&=&
\sum_q
\frac{4\pi\alpha^2}{9q^2} \hat e^2_q
\int d\vec{p}_q \int d\vec{p}_{\bar{q}}
\int  d\vec{p}_N \rho_{p_d}(\vec{p}_N)
f^{q}_{\vec p_p}(\vec{p}_q)f^{\bar{q}}_{\vec p_N}(\vec{p}_{\bar{q}})
\delta\left(q^2-(p_q+p_{\bar{q}})^2\right) .
\nonumber\\
\label{eq:pd-dy-01}
\end{eqnarray}
The above expression is for the contribution from an
anti-quark  in the nucleon $N$ of the deuteron and a quark in the 
projectile proton. Other contributions have the similar expressions,
just  with different quark indices.

By using the definitions of the
parton distributions,  Eq.~(\ref{eq:pdf-1})
for $f^{\bar{q}}_{N}(x^N_2)$ and  Eq.~(\ref{eq:pdf-2}) 
for $f^{q}_{p}(x_1)$, 
Eq.~(\ref{eq:pd-dy-01})
 can be written in terms of momentum fraction variable
$x_1$ for $q$ in the projectile proton $p$ and $x^N_2$ of the nucleon
$N$ in the deuteron. We then obtain
\begin{eqnarray}
\frac{d\sigma^{pd}(p_p,p_d)}{dq^2}
&=&
\sum_q \frac{4\pi\alpha^2}{9q^2} \hat e^2_q
\int dx_1\int dx^N_2\int d\vec{p}_{N}                                
\rho_{p_d}(\vec{p}_{N})
f^{q}_{p}(x_1)f^{\bar{q}}_{N}(x^N_2)
\delta\left(q^2-(p_q+p_{\bar{q}})^2\right) .
\nonumber\\
\label{eq:dy-pd-00}
\end{eqnarray}

\begin{figure}[t]
\includegraphics[clip,width=0.75\textwidth]{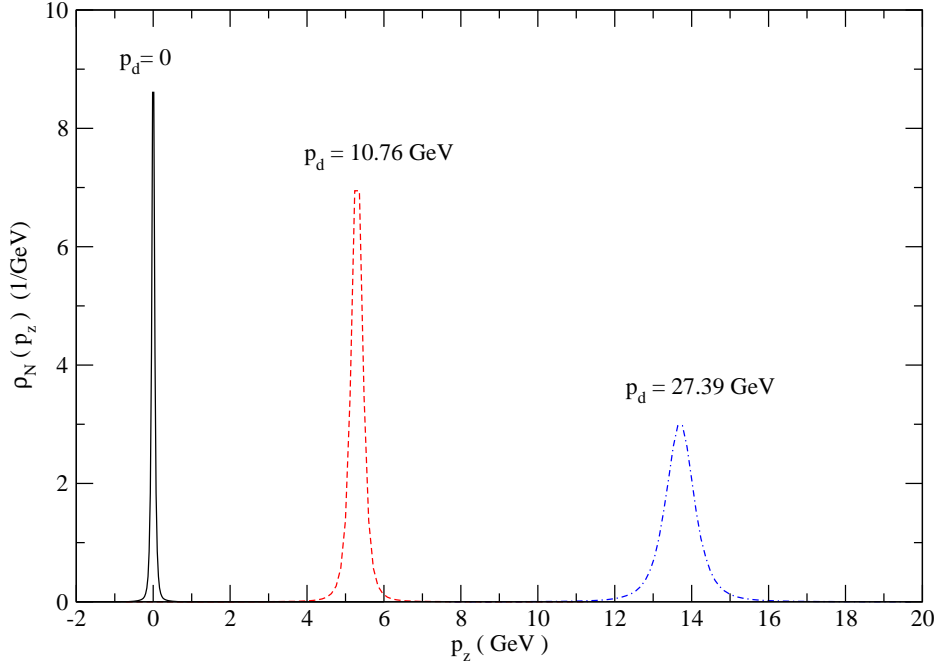}
\caption{(color online)
Nucleon momentum distribution $\rho_{p_d}(p_z) =\int d\vec p_\perp
\rho_{p_d}(p_z,\vec p_\perp)$ in a deuteron
moving with momentum $p_d$ in the $z$-direction.
The deuteron wave function of Ref.~\cite{v14} is used.}
 \label{fig:rho-v14}
\end{figure}

Similar to the $pp$ case, the deuteron momentum
is chosen to be in the $z$-direction: 
$\vec{p}_d =(p^z_d, \vec{p}_{d\perp}=\vec 0)$.  
Before we proceed further, it is necessary to relate the momentum fraction
$x^N_2$ in Eq.~(\ref{eq:dy-pd-00}) to $x_2$
which is determined by the experimental variables $M$, $s$ and $x_f$ through
the relations: $x_1x_2=M^2/s$ and
$x_f=x_1-x_2 = p^\gamma_\parallel/(\sqrt{s}/2)$.
Since our derivation is based on the impulse approximation that the
parton  is emitted from the nucleon in the deuteron, it is
appropriate to assume that the momentum of the emitted
parton is $p_{\bar{q}}^z=x_2 p^z_{\text{ave}}$, where
$p^z_{\text{ave}}$ is the averaged nucleon momentum in the
deuteron defined by 
\begin{eqnarray}
(p^z_{\text{ave}})^2= \frac{\int \rho_{p_d}(\vec{p}) 
(p^z)^2 d\vec{p}}
{\int \rho_{p_d}(\vec{p}) d\vec{p}}
\label{eq:p-average}
\end{eqnarray}
Note that $\rho_{p_d}(\vec p)$ in the above equation is
calculated from the nucleon momentum distribution
$\rho_N(\vec p)$ [Eq.~(\ref{eq:deut-rhop})] in the deuteron rest frame
by using the relation Eq.~(\ref{eq:lt-dent}).
In Fig.~\ref{fig:rho-v14}, we show the dependence of the calculated 
$\rho_{p_d}(p^z) \equiv \int d\vec p_\perp \rho_{p_d}(p^z,\vec p_\perp)$ on a deuteron
momentum $p_d$.
As expected, we find that that $p^z_{\text{ave}} \sim p^z_d/2$ at each deuteron momentum.

Changing the integration variable by 
$x_2^N=p^z_{\bar{q}}/p^z_N=x_2p^z_{\text{ave}}/p^z_N$, we can write 
Eq.~(\ref{eq:dy-pd-00}) as
\begin{eqnarray}
\frac{d\sigma^{pd}(p_p,p_d)}{dq^2}
&=&
\frac{4\pi\alpha^2}{9q^2}\hat e^2_q
\int dx_1\int dx_2\int d\vec{p}_{N}
\rho_{p_d}(\vec{p}_{N})\frac{p^z_{\text{ave}}}{p^z_N}
f^{q}_{p}(x_1)f^{\bar{q}}_{N}(x_2p^z_{\text{ave}}/p^z_N)
\nonumber \\
&&\times 
\delta\left(q^2-(p_q+p_{\bar{q}})^2\right) .
\end{eqnarray}
Integrating over $q^2$ on both sides of the above equation, 
we then obtain an expression of the cross section in terms
of $x_1$ and $x_2$ which are defined by experimental kinematics
\begin{eqnarray}
\frac{d\sigma^{pd}(p_p,p_d)}{dx_1dx_2}
&=& 
\frac{4\pi\alpha^2}{9q^2}\hat e^2_q 
f^{q}_{p}(x_1) F^{\bar{q}}_{p_d,N}(x_2) ,
\label{eq:dy-pd-x1x2}
\end{eqnarray}
where the $\bar{q}$ contribution is isolated in
\begin{eqnarray}
F^{\bar{q}}_{p_d,N}(x_2) 
&=&
\int d\vec{p}_{N} \rho_{p_d}(\vec{p}_{N})
\frac{p^z_{\text{ave}}}{p^z_N} f^{\bar{q}}_{N}(x_2p^z_{\text{ave}}/p^z_N)  .
\label{eq:pdf-ave0}
\end{eqnarray}
Within the parton model, we should only keep the contribution
from $f^{\bar{q}}_{N}(x_2p^z_{\text{ave}}/p^z_N)$ with $x_2p^z_{\text{ave}}/p^z_N \leq 1$. 
The above equation can then be written as
\begin{eqnarray}
F^{\bar{q}}_{p_d,N}(x_2)
&=&
\int^\infty_{(x_2p^z_{\text{ave}})}dp^z_N
\frac{p^z_{\text{ave}}}{p^z_N} 
f^{\bar{q}}_{N}(x_2p^z_{\text{ave}}/p^z_N) \rho_{p_d}(p^z_N) ,
\label{eq:pdf-ave}
\end{eqnarray}
with
\begin{eqnarray}
\rho_{p_d}(p^z_N) 
&=& 
\int d\vec p_{N\perp} \rho_{p_d}(p^z_N,\vec p_{N\perp}) .
\label{eq:rhoz}
\end{eqnarray}

The derivation of Eq.~(\ref{eq:pdf-ave}) can be extended to have $q$ 
in deuteron and $\bar{q}$ in the projectile proton.
We finally obtain
\begin{eqnarray}
\frac{d\sigma^{pd}(p_p,p_d)}{dx_1dx_2}
&=&
\frac{4\pi\alpha^2}{9q^2}
\sum_{q}
\hat e^2_q 
\left[
f^{q}_{p}(x_1)
\sum_{N=p,n}F^{\bar{q}}_{p_d,N}(x_2)+ 
f^{\bar{q}}_{p}(x_1)\sum_{N=p,n} F^{q}_{p_d,N}(x_2)
\right] .
\label{eq:dy-pd-c}
\end{eqnarray}
We use the charge symmetry to calculate PDFs
for the neutron from that of proton:
$f^d_n =f^u_p$, $f^u_n =f^d_p$, $f^{\bar{d}}_n =f^{\bar{u}}_p$, $f^{\bar{u}}_n =f^{\bar{d}}_p$.
Furthermore $\rho_{p_d}(p^z_N)$ is the same for neutron and proton.
Including the charges for $u$ and $d$ quarks appropriately,
Eq.~(\ref{eq:dy-pd-c}) can be written as
\begin{eqnarray}
\frac{d\sigma^{pd}(p_p,p_d)}{dx_1dx_2}
&=&
\frac{4\pi\alpha^2}{9q^2} 
\left\{
\left[ \frac{4}{9}f^u_p(x_1)+\frac{1}{9}f^d_p(x_1) \right]
\left[ F^{\bar{u}}_{p_d,p}(x_2)+F^{\bar{d}}_{p_d,p}(x_2) \right] 
\right.
\nonumber \\
&& 
\quad
\qquad
\left. +
\left[ \frac{4}{9}f^{\bar{u}}_p(x_1)+ \frac{1}{9}f^{\bar{d}}_p(x_1) \right]
\left[ F^{u}_{p_d,p}(x_2) +F^{d}_{p_d,p}(x_2) \right]
\right\} .
\label{eq:dy-pd-cc}
\end{eqnarray}

The formula for calculating the contribution from pion momentum distribution can be
derived by the similar procedure.
We obtain 
\begin{eqnarray}
\frac{d\sigma^{pd}_\pi(p_p,p_d)}{dx_1dx_2}
&=&
\frac{4\pi\alpha^2}{9q^2}
\left[
\frac{4}{9}f^u_p(x_1)F^{\bar{u}}_{p_d,\pi}(x_2)+\frac{1}{9}f^d_p(x_1)F^{\bar{d}}_{p_d,\pi}(x_2)
\right.
\nonumber \\
&&
\quad\qquad
\left.
+\frac{4}{9}f^{\bar{u}}_p(x_1)F^{u}_{p_d,\pi}(x_2)+\frac{1}{9}f^{\bar{d}}_p(x_1)F^{d}_{p_d,\pi}(x_2)
\right] ,
\label{eq:dy-pd-cc-pi}
\end{eqnarray}
where $f^{{q}}_{k_\pi}(x)$ is PDFs for the pion
taken from Ref.~\cite{holt2005}, and the convolution function for the pion is 
\begin{eqnarray}
F^{{q}}_{p_d,\pi}(x_2)
&=&
\int^\infty_{(x_2k^z_{\text{ave}})}dk^z_\pi
\frac{k^z_{\text{ave}}}{k^z_\pi}
f^{{q}}_{\pi}(x_2k^z_{\text{ave}}/k^z_\pi) \rho_{p_d}(k^z_\pi) ,
\label{eq:pdf-ave-pi}
\end{eqnarray}
with
\begin{eqnarray}
\rho_{p_d}(k^z_\pi)
&=&
\int d\vec k_{\pi\perp} \rho_{p_d}(k^z_\pi,\vec k_{\pi\perp}) .
\label{eq:rhoz-pi} 
\end{eqnarray}
 Here the average pion momentum is defined by
\begin{eqnarray}
(k^z_{\text{ave}})^2&=& \frac{\int \rho_{p_d}(\vec{k}_\pi)
(k^z_\pi)^2 d\vec{k}_\pi}
{\int \rho_{p_d}(\vec{k}_\pi) d\vec{k}_\pi} .
\label{eq:pi-average}
\end{eqnarray}
 The pion momentum distribution
$\rho_{p_d}(\vec{k}_{\pi})$ in the above equations
is calculated 
from $\rho^{\text{exc}}_\pi(\vec{k})$ of Eq.~(\ref{eq:pi-dis}) by using 
the relation Eq.~(\ref{eq:lt-dent}).

\section{Numerical Results}
\label{sec:results}

As discussed in Ref.~\cite{peng2001}, the ratio $\bar{d}/\bar{u}$ in the
proton can be extracted from the data of the ratios between 
the $pd$ and $pp$ DY cross sections:
\begin{eqnarray}
R_{pd/pp}&=&
\frac{d\sigma^{pd}(p,p_d)}{dx_1dx_2}\bigg/
\left(2\frac{d\sigma^{pp}(p,p_p)}{dx_1dx_2}\right) \,, \label{eq:pd-ratio}
\end{eqnarray}
where $x_1$ and $x_2$ have been defined in Sec.~\ref{sec:numerical-1}.
We are interested in the effects of pion-exchange and nucleon Fermi motion
on this ratio.
The $pp$ cross section
$d\sigma^{pp}(p,p_p)/(dx_1dx_2)$ can be calculated
 from Eq.~(\ref{eq:dy-pp-cc}).
The $pd$  cross section ${d\sigma^{pd}(p,p_d)}/(dx_1dx_2)$ 
is the sum of the nucleon contribution calculated 
from Eq.~(\ref{eq:dy-pd-cc}) and the pion contribution from
Eq.~(\ref{eq:dy-pd-cc-pi}).
To compare with the results of  Ref.~\cite{peng2001}, the
nucleon PDFs $f^q_p(x)$ of CETEQ5m~\cite{ceteq5m} is used in
our calculations.
The PDFs $f^q_\pi(x)$ for the pion is taken from Ref.~\cite{holt2005}.

From Eq.~(\ref{eq:dy-pd-cc}), it is clear that the nucleon Fermi motion
effects are in $F^{q}_{p_d,N}(x_2)$ defined by in Eq.~(\ref{eq:pdf-ave}).
If we set $ p^z_{\text{ave}}/p^z_N \to 1$ in Eq.~(\ref{eq:pdf-ave}), 
$F^{q}_{p_d,N}(x_2) \to f^q_N(x_2)$ since $\int d\vec{p}_N \rho_{p_d}(\vec{p}_N)=1$ 
as defined by the normalization of states.
The calculation of Eq.~(\ref{eq:dy-pd-cc}) with $F^{q}_{p_d,N}(x_2) \to f^q_N(x_2)$
is then identical to that based on Eq.~(\ref{eq:dy-pd-exp}) of Ref.~\cite{peng2001}.
The differences between this calculation and that from using 
Eqs.~(\ref{eq:dy-pd-cc}) and~(\ref{eq:pdf-ave}) will indicate the importance of 
nucleon Fermi motion effect on $pd$ DY cross sections.

\begin{figure}[t]
\includegraphics[width=0.75\textwidth,clip]{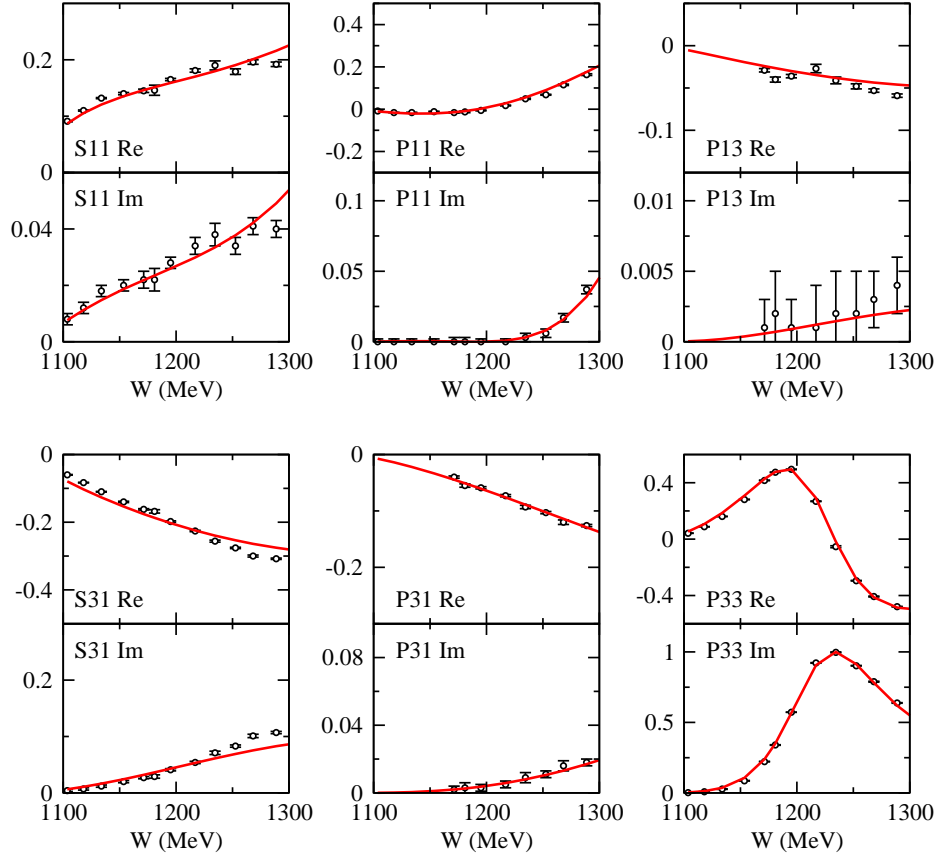}
\caption{\label{fig:pn-amp}
(color online)
Results of the fit to $\pi N$ scattering amplitudes~\cite{said} up to $W=1.3$ GeV.
}
\end{figure}

To calculate the pion contribution with Eq.~(\ref{eq:dy-pd-cc-pi}), 
we need to first evaluate $F^q_{p_d,\pi}(x_2)$ defined by Eq.~(\ref{eq:pdf-ave-pi}).
The pion momentum $\rho_{p_d}(\vec{k}_\pi)$ in Eq.~(\ref{eq:pdf-ave-pi})
is calculated from using the relation Eq.~(\ref{eq:lt-dent})
and $\rho^{\text{exc}}_\pi(\vec{k})$ defined by Eq.~(\ref{eq:pi-dis}).
We see from  Eq.~(\ref{eq:pi-dis}) that the pion momentum distribution 
$\rho^{\text{exc}}_\pi(\vec{k})$ depends on the $\pi NN$ form factor 
$F(\Lambda_{\pi NN},\vec{k})$ [Eq.~(\ref{eq:v-f})].
Following the previous $\pi NN$ studies~\cite{garcilazo,lee-nn},
this form factor must be consistent with $\pi N$ scattering data.
In this work, we apply the $\pi N$ model formulated  in Ref.~\cite{knls10} 
to determine the $F(\Lambda_{\pi NN},k)$ by fitting 
the $\pi N$ partial wave amplitudes~\cite{said} up to invariant mass $W=1.3$ GeV. 
The $\pi N$ scattering within this model has been given in Ref.~\cite{knls10} 
and will not be repeated here.
Our fits are shown in Fig.~\ref{fig:pn-amp}. The resulting parameters are not
relevant to this work and are therefore not  presented.
For our calculation, we only need  the resulting $\pi NN$ form factor.

\begin{figure}[t]
\includegraphics[width=0.75\textwidth,clip]{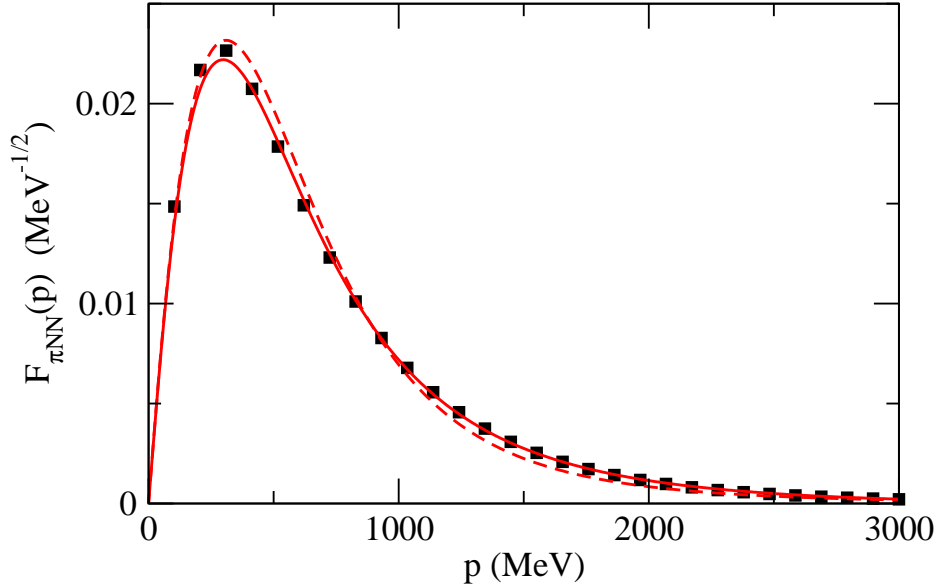}
\caption{\label{fig:pnn-ff}
(color online)
Filled squares are the $\pi NN$ form factor obtained from fitting the $\pi N$ 
partial wave amplitudes~\cite{said}.
Solid and dashed curves are two different fits to  the form factor
using the parametrization of Eqs.~(\ref{eq:ff-2}) and dipole form 
$F(\Lambda_{\pi NN}, k) = \left(\frac{\Lambda^2_{\pi NN}-\kappa^2}{\Lambda^2_{\pi NN}+k^2}\right)^2$
with $\Lambda_{\pi NN} = 810.6$ MeV}
\end{figure}

We see in Fig.~\ref{fig:pnn-ff} that the resulting $\pi NN$ form factor
can be fitted by the following modified dipole form
\begin{equation}
F(\Lambda_{\pi NN}, k) = \left(\frac{\Lambda^2_{\pi NN}-\kappa^2}{\Lambda^2_{\pi NN}+k^2}\right)^2
\left[1+a(1+k^2/\kappa^2)\right]\exp\left[-b (1+k^2/\kappa^2)\right],
\label{eq:ff-2}
\end{equation}
where $\kappa = m_\pi\sqrt{1-[m_\pi^2/(4m_N^2)]}$,
$\Lambda_{\pi NN}= 685.7$ MeV, $a=1.67\times 10^{-3}$, $b=2.79\times 10^{-4}$. 
It is close to the usual dipole form 
$F(\Lambda_{\pi NN}, k) = \left(\frac{\Lambda^2_{\pi NN}-\kappa^2}{\Lambda^2_{\pi NN}+k^2}\right)^2$
with $\Lambda_{\pi NN} = 810.6$ MeV.

\begin{figure}[t]
\includegraphics[clip,width=0.75\textwidth]{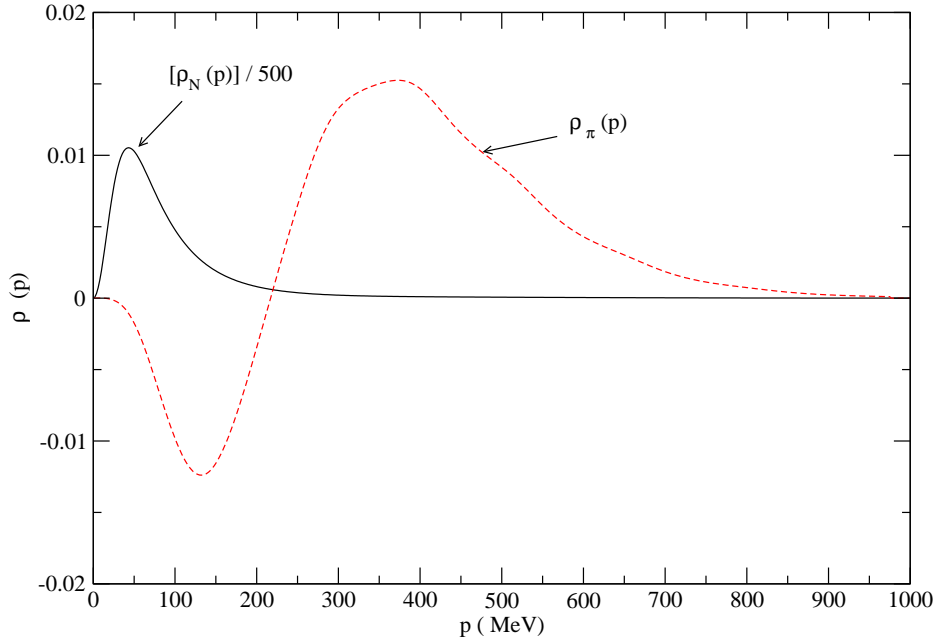}
\caption{ 
(color online)
The momentum distribution $4\pi p^2\rho(p)$ of 
the pion ($\pi$) and the nucleon ($N$) in the deuteron. 
Note that $4\pi p^2\rho_N(p)$ is multiplied by a factor $1/500$. }
\label{fig:rho-n-p}
\end{figure}

The pion momentum distribution 
$\rho^{\text{exc}}_\pi(\vec{k})$ calculated from Eq.~(\ref{eq:pi-dis})
with the $\pi NN$ form factor given in Eq.~(\ref{eq:ff-2}) is
the dashed curve in Fig.~\ref{fig:rho-n-p}.
Here we also show the nucleon momentum distribution $\rho_N(p)$ (solid curve).
Note that $\rho^{\text{exc}}_\pi(p)$ changes sign at $p \sim 200$ MeV. 
This sign change is also seen in the calculation of pion-excess in Ref.~\cite{fpw}, except that
their magnitudes are much larger because they use a much larger $\pi NN$ cutoff $\Lambda \sim 1400$
MeV for a dipole form of a non-relativistic $NN$ potential.

\begin{figure}[t]
\includegraphics[clip,width=0.75\textwidth]{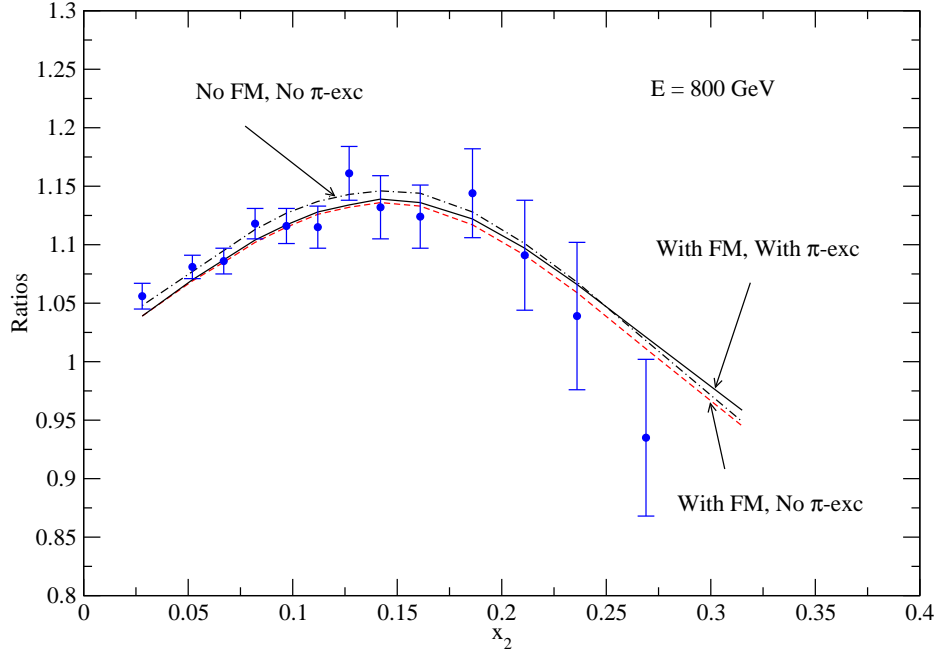}
\caption{
(color online)
Ratio $R_{pd/pp}$ at $E=800$ GeV. Data are from Ref.~\cite{peng2001}. 
With (No) FM denotes that Fermi motion is included (not included). 
With (No) $\pi$-exc denotes that pion-exchange is included (not included).
Note that $x_1$ for each $x_2$ is determined by Eq.~(\ref{eq:xf}) 
and given in Ref.~\cite{peng2001}.
}
\label{fig:ratiopi-800-c850}
\end{figure}

With the input specified above, we can calculate ratio
$R_{pd/pp}$ defined by Eq.~(\ref{eq:pd-ratio}). 
We compare three results:
(1) No nucleon Fermi motion (FM) and no pion-exchange ($\pi$-exc) from
using Eq.~(\ref{eq:dy-pd-cc}) with $F^{q}_{p_d,N}(x_2) \to f^q_N(x_2)$;
(2) With FM and no $\pi$-exc from using Eq.~(\ref{eq:dy-pd-cc});
(3) With FM and with $\pi$-exc from adding the results from using
Eq.~(\ref{eq:dy-pd-cc}) and Eq.~(\ref{eq:dy-pd-cc-pi}).
 
In Fig.~\ref{fig:ratiopi-800-c850},
the calculated $R_{pd/pp}$  at 800 GeV are compared with the data 
of Ref.~\cite{peng2001}.
Our results with no Fermi motion 
and no pion-exchange (dot-dashed curve) are similar to that presented
in Ref.~\cite{peng2001}.
The differences between the dash-dotted  and dashed curves are due to the Fermi motion of
nucleon inside the deuteron. 
The solid curve also include the pion-exchange effects. All three results
are close to the data.
Clearly, 
the nucleon Fermi motion and pion-exchange effects are small in the 
region covered by this experiment.
Our results shown in Fig.~\ref{fig:ratiopi-800-c850} suggest
 that the simple formula Eq.~(\ref{eq:dy-pd-exp}) is valid to extract the
$\bar{d}/\bar{u}$ ratio in the proton in the small $x_2 \lesssim 0.3$ region.

\begin{figure}[t]
\includegraphics[clip,width=0.75\textwidth]{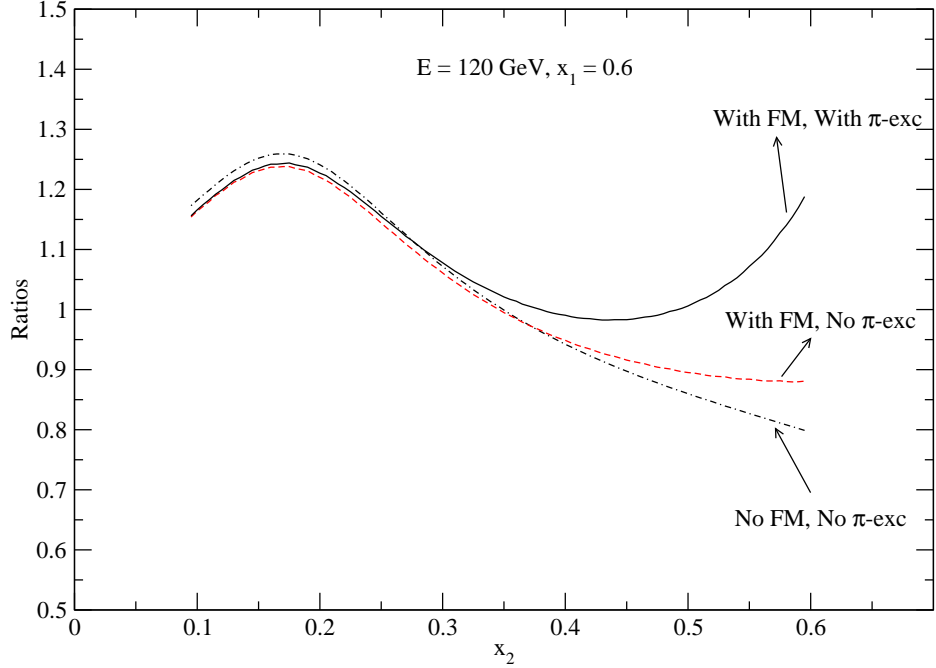}
\caption{
(color online)
Ratio $R_{pd/pp}$ at $E=120$ GeV and $x_1 = 0.6$.
With (No) FM denotes that Fermi motion is included (not included). 
With (No) $\pi$-exc denotes that pion-exchange is included (not included).  
}
\label{fig:ratiopi-120-c850}
\end{figure}

\begin{figure}[t]
\includegraphics[clip,width=0.75\textwidth]{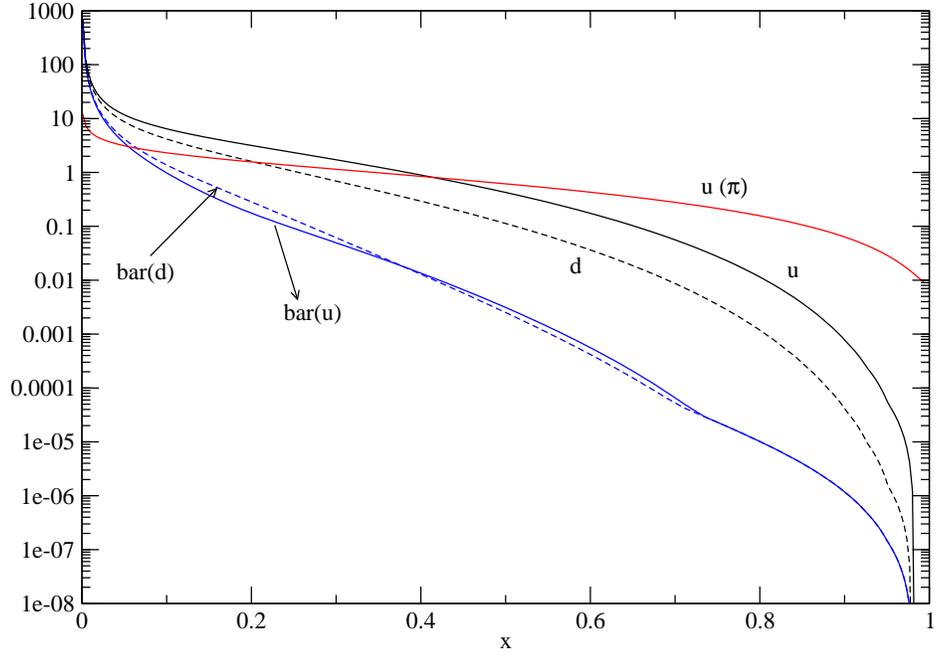}
\caption{ 
(color online)
Comparison of PDFs of the proton~\cite{ceteq5m} 
[$\bar{u}$ is denoted by bar(u) and $\bar{d}$  by bar(d)]
with that of the pion~\cite{holt2005} [$u(\pi)$, note that $\bar{u}(\pi) = u(\pi)$]. 
}
\label{fig:ceteq5-pdf}
\end{figure}

To facilitate the analysis of the forthcoming data from Fermilab,
we present our prediction at 120 GeV
in Fig.~\ref{fig:ratiopi-120-c850}.
We see that the Fermi motion  and pion-exchange effects are small
 in the $x_2 < 0.4$. However these two effects are significant at 
larger $x_2$.
We have observed that the rapidly raising
effect due to pion-exchange is due to the fact that the
parton distribution in the pion is much larger than that for
the nucleon at large $x$, as seen in Fig.~\ref{fig:ceteq5-pdf}.
Clearly, it is necessary to include the Fermi motion  
and pion-exchange effects to extract the ratio $\bar{d}/\bar{u}$ in
the proton from the data of $R_{pd/pp}$ in the  large $x_2$ region.

\section{summary}
\label{sec:summary}

For investigating the pion-exchange and nucleon Fermi-motion effects on
the DY process in proton-deuteron($pd$) reactions, 
we have derived convolution formula starting with a nuclear model 
within which the deuteron has $NN$ and $\pi NN$ components.
The nucleon Fermi motion is included by the convolution of PDFs of the nucleon over
the nucleon momentum distribution calculated from the $NN$ component.
The contribution from the $\pi NN$ component is expressed in terms of 
a convolution of PDFs of the pion over a pion momentum distribution that 
depends sensitively on the $\pi NN$ form factor.
With a $\pi NN$ form factor determined by fitting the $\pi N$ scattering
data up to invariant mass $W=$ 1.3 GeV, we find that the pion-exchange and 
nucleon Fermi-motion effects can change significantly the ratios between 
the proton-deuteron and proton-proton DY cross sections
$R_{pd/pp}=\sigma^{pd}/(2\sigma^{pp})$ in the region where the partons emitted 
from the target deuteron are in the Bjorken $x_2 \gtrsim 0.4$ region.
The calculated ratios $R_{pd/pp}$ at 800 GeV agree with the available data.
For analyzing the forthcoming data from Fermilab, we also have made
predictions at 120 GeV.

\begin{acknowledgments}
We would like to thank Donald Geesaman, Roy Holt, and Jen-Chieh Peng 
for their very helpful discussions.
This work is supported by the U.S. Department of Energy, Office of Nuclear Physics Division,
under Contract No. DE-AC02-06CH11357.
HK acknowledges the support by the HPCI Strategic Program (Field 5
``The Origin of Matter and the Universe'') of Ministry of Education (Japan).
This research used resources of the National Energy Research Scientific Computing Center,
which is supported by the Office of Science of the U.S. Department of Energy
under Contract No. DE-AC02-05CH11231, and resources provided on ``Fusion,''
a 320-node computing cluster operated by the Laboratory Computing Resource Center
at Argonne National Laboratory.
\end{acknowledgments}

\end{document}